\documentclass[fleqn,usenatbib]{mnras}

\usepackage{newtxtext,newtxmath}

\usepackage[T1]{fontenc}

\DeclareRobustCommand{\VAN}[3]{#2}
\let\VANthebibliography\thebibliography
\def\thebibliography{\DeclareRobustCommand{\VAN}[3]{##3}\VANthebibliography}

\usepackage{graphicx}	
\usepackage{amsmath}	

\title[The passive evolution of the most massive quiescent galaxies]{Growing in number, passive in nature: tracing the evolution of the most massive quiescent galaxies since $z \sim 0.8$ with BOSS and DESI}

\author[F. R. Ditrani et al.]{
F. R. Ditrani,$^{1,2}$\thanks{E-mail: f.ditrani1@campus.unimib.it}
M. Fossati,$^{1,2}$
M. Longhetti,$^{2}$
F. La Barbera,$^{3}$
A. Iovino,$^{2}$
C. Maraston,$^{4}$
D. Thomas$^{5,6}$,
\and and D. Bevacqua$^{7}$
\\
$^{1}$Università degli studi di Milano-Bicocca, Piazza della scienza, I-20125 Milano, Italy\\
$^{2}$INAF-Osservatorio Astronomico di Brera, via Brera 28, I-20121 Milano, Italy\\
$^{3}$INAF - Osservatorio Astronomico di Capodimonte, Via Moiariello 16, I-80131 Napoli\\
$^{4}$Institute of Cosmology, University of Portsmouth, Burnaby Road, Portsmouth PO1 3FX, UK\\
$^{5}$School of Mathematics and Physics, University of Portsmouth, Lion Gate Building, Portsmouth PO1 3HF, UK\\
$^{6}$School of Physics and Astronomy, University of Leeds, Leeds, LS2 9JT, UK\\
$^{7}$ INAF - Osservatorio Astronomico di Roma, Via Frascati 33, 00078 Monte Porzio Catone, Italy
}

\date{Accepted XXX. Received YYY; in original form ZZZ}

\pubyear{\the\year{}}

\begin{document}
\label{firstpage}
\pagerange{\pageref{firstpage}--\pageref{lastpage}}
\maketitle

\begin{abstract}
Luminous Red Galaxies (LRGs) are among the most massive galaxies at any epoch, and lack ongoing star formation. As systems hosting most of the baryonic mass in the local Universe, they preserve imprints of the quenching mechanisms in the early Universe.
We exploited the large BOSS and DESI spectroscopic datasets to perform the first homogeneous and continuous mapping of the evolution of stellar population properties of a complete sample of the most massive LRGs ($\log (M_*/\mathrm{M_\odot})> 11.5$) at $0.15 < z < 0.8$. By consistently fitting the same spectral indices at all redshifts, we measured trends of [Fe/H], [$\alpha$/Fe], and light–weighted age as a function of redshift. These galaxies exhibit a passive light-weighted age evolution and flat [Fe/H] and [$\alpha$/Fe] trends towards lower redshift, indicating genuinely passive evolution. These trends are robust against the choice of stellar population models and analysis assumptions, and they support the predictions from IllustrisTNG, which predict negligible chemical evolution for the most massive quenched systems at $z \le 0.8$. 
Our results suggest that, despite nearly $5$ Gyr of cosmic time and a $3-4\times$ increase in number density, the stellar population properties of massive quiescent galaxies remain essentially unchanged since $z \sim 0.8$. This shows a negligible progenitor bias below $z \sim 0.8$, and a genuinely passive evolution. Newly added systems after $z \sim 0.8$ were already largely quenched and chemically mature, while subsequent evolution was dominated by dry mergers without altering the bulk of the stellar populations.
\end{abstract}




\section{Introduction}

Luminous red galaxies (LRGs) are a population of massive, quiescent galaxies identified by their high luminosities and red rest-frame optical colours, indicative of a dominant old stellar populations. LRGs rank among the most massive galaxies at their epoch ($\log (M_*/\mathrm{M_\odot})> 11.5$), with no ongoing star formation \citep[e.g.][]{eisenstein2001,barber2007}. 
The LRGs are highly biased tracers of the overall cosmic matter distribution and thus ideal to study large scale structures. As the dominant systems hosting the bulk of the baryonic mass in the local Universe \citep[][and references therein]{renzini2006}, these massive, quiescent galaxies preserve clear fingerprints of the physical processes that quenched star formation in the early Universe, serving as cosmic laboratories for understanding the formation and evolution of massive galaxies \citep[][]{gallazzi2005ages,thomas2005epochs}.
A powerful way to explore the physical mechanisms of LRGs assembly is by tracing their stellar population properties. Tracing their stellar ages, total metallicities, and [$\alpha$/Fe] abundance ratios offer powerful diagnostics of their star formation histories. The [$\alpha$/Fe] ratio, and in particular [Mg/Fe], is sensitive to the duration of star formation \citep{tinsley1979,thomas1999} because $\alpha$-elements (such as Mg) are predominantly produced by core-collapse supernovae from massive stars, which explode on short timescales ($\sim 10$ Myr). In contrast, the bulk of iron is released later by Type Ia supernovae, which arise from longer-lived progenitors and enrich the interstellar medium on timescales of $\sim 1$ Gyr \citep{greggio1983,jafariyazani2025}. As a result, under the assumption of a constant IMF \citep{ferreras2015,fontanot2018} a high [Mg/Fe] indicates a short, intense burst of star formation completed before significant iron enrichment could occur \citep{matteucci1987}, while lower values point to more extended star formation histories. Combining [Mg/Fe] with age and metallicity estimates allows us to reconstruct both the timing and duration of star formation episodes, thus providing a physically grounded view of galaxy assembly.
First statistical studies based on the Sloan Digital Sky Survey \citep[SDSS,][]{eisenstein2001,eisenstein2005} have played a key role in understanding the properties of LRGs in the local Universe up to $z < 0.1$. In the local Universe, LRGs exhibit old stellar populations with high stellar metallicity and enhanced $\alpha$-element abundances relative to iron \citep{thomas2005epochs,barber2007,thomas2010}. Going up to $z \sim 0.4$, the Baryon Oscillation Spectroscopic Survey \citep[BOSS,][]{dawson2013,reid2016} confirmed the properties of LRGs, indicating short and intense star formation episodes at high redshift.
The SDSS and BOSS surveys have therefore established a robust observational framework, highlighting the early assembly and passive evolution of LRGs \citep{maraston2013,bundy2017}. However, these surveys are complete up to $z \sim 0.4$ and in a narrow higher redshift slice \citep[$0.55<z<0.6$,][]{leauthaud2016} for LRGs in the very high mass end ($\log (M_*/\mathrm{M_\odot}) > 11.6$), leaving open key questions about how their stellar populations evolved at earlier epochs.

Intermediate redshifts ($ 0.4 \le z \le 1$) represent a crucial stage in the evolution of LRGs, as at $z \sim 1$ the fraction of quiescent galaxies with $\log (M_*/\mathrm{M_\odot}) > 11.5$ reaches its maximum and increases in number density only slowly at lower redshift \citep{muzzin2013}. This epoch thus marks the point when the majority of the most massive galaxies have completed their stellar mass assembly and transitioned onto the passive sequence. Studying this redshift range is therefore key to understanding how and when massive galaxies settle into the quiescent population observed locally.

Apart from a few smaller studies targeting intermediate redshift galaxies \citep[e.g.][]{gallazzi2014, choi2014assembly}, one of the first large statistical spectroscopic surveys in this redshift regime has been the Large Early Galaxy Astrophysics Census \citep[LEGA-C,][]{van2016vlt,van2021large}. LEGA-C allowed detailed measurements of the stellar population properties of massive quiescent galaxies at these intermediate redshifts \citep[e.g.][]{cappellari2023,ditrani2025}, with a few studies concerning the [$\alpha$/Fe] abundances \citep[e.g.][]{beverage2021,beverage2023,bevacqua2023}.
However, LEGA-C targets a single field \citep[COSMOS,][]{scoville2007} covering only $\sim 1.6~\mathrm{deg}^2$, and given the low number density of the most massive systems ($\log (M_*/\mathrm{M_\odot}) > 11.5$), the resulting sample contains too few massive quiescent galaxies to enable a statistically robust analysis of their properties. A wider-area spectroscopic survey is therefore required to probe the high-mass end of the quiescent population with sufficient statistics.

Recently, the Dark Energy Spectroscopic Instrument \citep[DESI,][]{desi2016a,desi2016b} is undertaking the largest LRGs survey to date. Indeed, it will observe around $8$ million LRGs in the redshift range $0.4 < z < 1$, covering around $14000$ deg$^2$. Compared to observations from the surveys described above, DESI provides a significantly higher sampling density and extends the coverage to higher redshifts. In particular, the LRGs sample achieves a comoving number density of $5*10^{-4} h^3 Mpc^{-3}$, obtaining a complete population of massive quiescent galaxies with $\log (M_*/\mathrm{M_\odot}) > 11.5$ within $0.4 \le z \le1$ \citep[][]{zhou2023}.

In this work, we exploited the extensive spectroscopic datasets from the BOSS and DESI surveys to carry out a comprehensive analysis of the evolution of stellar population properties of LRGs across cosmic time. For the first time, we derived ages, total metallicities and in particular [$\alpha$/Fe] abundance ratios for a large, mass-complete sample of massive LRGs that continuously spans the redshift range $0.15 < z < 0.8$. We then investigated how their stellar population properties evolve across the last $\sim 7$ Gyr of the cosmic time. The continuous redshift coverage of the joined BOSS and DESI sample represents a key advancement, enabling us to probe the evolutionary trends of massive quiescent galaxies in a homogeneous and consistent way throughout this wide redshift range.

This paper is structured as follows: in Sect~\ref{sec:datasel} we present the selected LRG data used in our analysis and describe the treatment of the spectra to obtain high-quality stacked spectra suitable for our study. In Sect~\ref{sec:analysis} we describe in detail the procedures adopted to retrieve the stellar population parameters
of our selected sample. In Sect.~\ref{sec:result} we present the results that we obtained and the discussion of our analysis.
Throughout the paper, we adopt a standard $\Lambda$CDM cosmology with $\Omega_M = 0.3089$, $\Omega_\Lambda = 0.6911$, and $H_0 = 67.8$ km s$^{-1}$ Mpc$^{-1}$ 
\citep{planck2016}. Magnitudes are in the AB system \citep{oke1974absolute}. 

\section{Data and sample selection} 
In this section we present the LRGs samples drawn from the DESI and BOSS surveys and their most important properties, including the stellar mass completeness limits and redshift distributions.

\label{sec:datasel}
\subsection{DESI}
We used the DESI Data Release 1 \citep[DR1,][]{desidr12025arXiv}, which provides all the data acquired during the first $13$ months of the DESI main survey. The DESI DR1 includes spectra for more than $18$ million unique targets, covering around $9500$ deg$^2$. Although the DESI LRGs survey will be completed at the end of the full observational campaign, the DR1 offers a representative and statistically significant dataset of the final sample. The DESI observations have a spectral resolution of R $= 2000-5000$ and cover the observed wavelength range $3600-9800 \ \text{\AA}$.
We selected LRGs within two mass bins [$11.3< \log (M_*/\mathrm{M_\odot}) < 11.5$, $\log (M_*/\mathrm{M_\odot}) > 11.5$], for which completeness is $\sim 80\%$ and $\sim 90\%$, respectively \citep{zhou2020,zhou2023}. With this selection, we obtained $300019$ galaxies in the lower mass bin and $100952$ for the higher one. We limited our selection to the redshift range $0.4 < z < 0.8$, which ensures that all spectral indices used in our analysis (see Sect.~\ref{subsec:fullindex}) are consistently available across the full sample.

\subsection{BOSS}
With the aim of extending the redshift range of our study, we complement the DESI dataset with the LRG sample from the BOSS survey \citep[][]{eisenstein2011,dawson2013}, whose selection is focussed on the redshift range $0.15<z<0.6$. The BOSS observations have a spectral resolution of R $= 1500-2500$ and cover the observed wavelength range $3600-10400 \ \text{\AA}$.
The BOSS LRGs sample consists of two subsamples: a low-redshift sample (LOWZ), that contains galaxies up to $z\approx 0.4$, and a high-redshift sample (CMASS) for galaxies at $z > 0.4$.
According to \cite{leauthaud2016}, the LOWZ selection is $80\%$ complete for LRGs with $\log (M_*/\mathrm{M_\odot}) > 11.6$ in the redshift range $0.15 <z < 0.43$, while the CMASS sample is $80\%$ complete only in the narrow redshift slice $0.51 < z < 0.61$ for the same mass range. We therefore consider only the LOWZ selection, considering a single mass bin [$\log (M_*/\mathrm{M_\odot})> 11.6$] in the redshift range $0.15 < z < 0.4$ (obtaining $9528$ galaxies), to ensure a fair comparison with the complete DESI sample and to study the evolutionary trends from $z = 0.8$ down to $z = 0.15$. 
In addition, we selected LRGs in the redshift range $0.35 < z < 0.6$ as a comparison sample of galaxies observed in both in DESI and BOSS surveys, in order to assess the consistency of these two datasets (App.~\ref{sec:appconsistency}).

\begin{figure}
   \centering
   \includegraphics[width=0.5\textwidth]{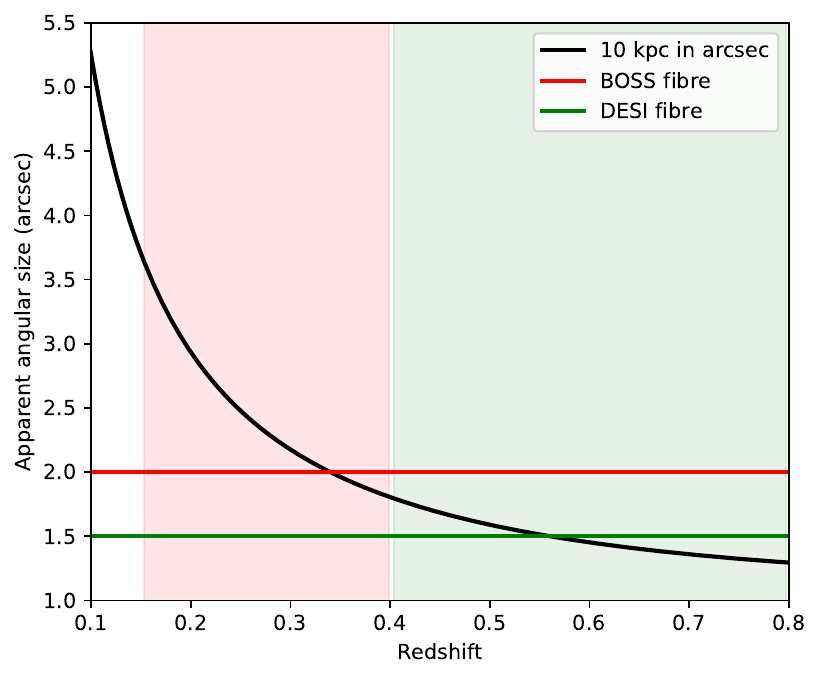}
      \caption{Apparent angular size (in arcsec) of a typical massive quiescent galaxy with a physical effective diameter of $10$ kpc (R$_\text{eff} = 5$ kpc) as a function of redshift. The horizontal red (green) line indicates the BOSS (DESI) fibre aperture. The shaded red (green) area marks the redshift range considered for the BOSS (DESI) sample.}
         \label{fig:effectiveradius}
   \end{figure}

When comparing BOSS and DESI data, aperture effects must be considered, as the two surveys employ fibres of different sizes. The DESI fibre has a diameter of $1.5$ arcsec, while the BOSS fibre is slightly larger, with a $2$ arcsec diameter. As shown in Fig.~\ref{fig:effectiveradius}, the DESI and BOSS fibre apertures correspond to roughly one effective radius ($\sim 5$ kpc) for a typical massive quiescent galaxy \citep[see][]{favole2018}. However, the apparent size changes rapidly for the redshifts covered by BOSS, in particular at $z < 0.3$, therefore we will take into account this difference when comparing the physical properties derived from the two surveys.

\begin{figure}
   \centering
   \includegraphics[width=0.5\textwidth]{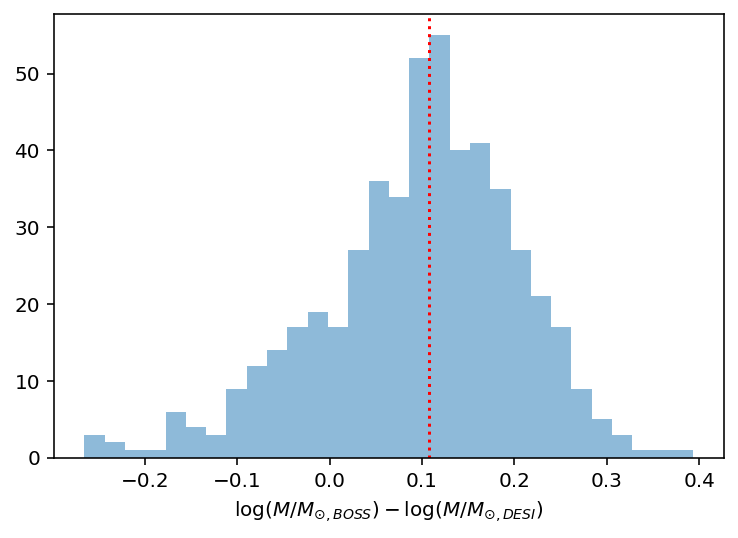}
      \caption{Histogram of the stellar mass differences for the matched galaxies in DESI and BOSS at $0.15 < z < 0.4$. The dotted red line marks the median value of the distribution.}
         \label{fig:confrmass_samegalaxies}
   \end{figure}

We investigated potential differences of the stellar mass estimates between the DESI and BOSS surveys, by selecting LRGs observed in both, at $0.15 < z < 0.4$, and applying the DESI mass threshold used for the sample comparison ($ \log (M_*/\mathrm{M_{\odot,DESI}}) > 11.5$), for which we obtained $517$ matched galaxies. Stellar masses in the BOSS sample were derived assuming a Kroupa IMF \citep[see][]{maraston2013}, while DESI masses were estimated assuming a Chabrier IMF. In addition, the two surveys rely on different photometric data and SED fitting procedures.  Figure~\ref{fig:confrmass_samegalaxies} shows the stellar mass differences for the matched galaxies in DESI and BOSS.  We found a median offset of $\sim 0.1$ dex, fully consistent with expected IMF-driven difference between Kroupa and Chabrier IMF assumptions \citep[$\sim 0.05$ dex, e.g.][]{longhetti2009,bernardi2010} and different templates assumptions in the SED fitting procedures. 
We verified this by applying the the median offset described above to the BOSS stellar masses of the main sample, and after this correction the two distributions, along with their median values ($\log (M_*/\mathrm{M_{\odot,DESI}}) = 11.57$, $ \log (M_*/\mathrm{M_{\odot,BOSS}}) = 11.56$), are fully consistent, showing consistent shapes.
Therefore, since DESI masses are smaller by $\sim 0.1$ dex than the BOSS ones, the $\log (M_*/\mathrm{M_\odot})> 11.5$ DESI sample corresponds to the selection of BOSS galaxies with $\log (M_*/\mathrm{M_\odot})> 11.6$.

\subsection{Further spectroscopic selection}
Although the LRG samples are dominated by quiescent galaxies, 
their selection on a purely photometric basis means that a fraction of emission-line galaxies is inevitably included. To obtain a clean sample of quiescent galaxies, whose absorption lines are unaffected, or only minimally affected, by ongoing star formation or nuclear activity, we applied an additional selection criterion. We use the following spectroscopic indicators: [OII]$_{\lambda3727}$, H$\delta$, H$\beta$, and [OIII]$_{\lambda5007}$, to define a subsample of spectroscopically confirmed quiescent galaxies satisfying these criteria:

\begin{itemize}
    \item EW([OIII]$_{\lambda5007}$) $< 1\,\text{\AA}$;
    \item EW(H$\beta$) $< 1\,\text{\AA}$;
    \item EW([OII]$_{\lambda3727}$) $< 5\,\text{\AA}$;
    \item EW(H$\delta$) $< 1\,\text{\AA}$.
\end{itemize}
where EW stands for Equivalent Width defined such that positive values correspond to emission. These criteria are designed to select galaxies with no detectable [OIII]$_{\lambda5007}$, H$\beta$, and H$\delta$ emission lines and at most a weak [OII]$_{\lambda3727}$ emission, which can occur in quiescent galaxies at these redshifts \citep[see][]{maseda2021,ditrani2025}. Regarding the DESI sample, we used the emission lines measured from the DR1/Iron value-added catalogue to follow the criteria described above, while for the BOSS sample we used the emissionLinesPort table from the SDSS database. Using the criteria described above, we obtained $224589$ galaxies from the DESI survey ($170901$ in the lower mass bin, $53688$ in the higher one), and $6160$ galaxies for the BOSS one.
Finally, we added a last criterion, EW(H$\delta$) $ > -3\,\text{\AA}$, to exclude post-starburst galaxies from our sample, following the definition of \cite{poggianti2009} and \cite{werle2022}. Galaxies with stronger H$\delta$ in absorption are recently quenched galaxies, which are likely to have experienced a different evolutionary track compared to the older quiescent galaxies that compose our sample. 
The final sample contains $194419$ quiescent galaxies from the DESI survey ($149243$ in the lower mass bin, $45176$ in the higher one), and $5954$ from the BOSS selection. Table~\ref{table:numbers} summarises the selected galaxies after each selection cut.

\begin{table}
\caption{Number of galaxies in the BOSS and DESI samples after all the selection cuts applied in this work.}
\centering  
\begingroup
\begin{tabular}{lll}
\hline\hline      
&BOSS sample & DESI sample  \\
\hline
LRG selection & 9528 & 400971 \\
Emission lines cut & 6160 & 224589 \\
Post-starburst cut & 5954 & 194419 \\
\hline

\end{tabular}
\endgroup
\label{table:numbers}
\end{table}

\subsection{Stacked spectra}
The BOSS and DESI samples provide us with a rich statistical sample of LRGs, albeit at a median SNR per $\text{\AA}$ insufficient to robustly determine the stellar population parameters of the selected population on individual spectra. We therefore stack blocks of individual spectra to boost the SNR  in order to reach an SNR of $\sim 20/\text{\AA}$ in the rest-frame range $4000-4200\,\text{\AA}$. In the DESI sample, at $0.4<z<0.8$, the median SNR of individual spectra in the considered wavelength range is around $4/\text{\AA}$. Therefore, we stacked the individual spectra in groups of $50$ objects, in bins of $\Delta z = 0.01$, within two mass bins [$11.3< \log (M_*/\mathrm{M_\odot}) < 11.5$, $\log (M_*/\mathrm{M_\odot}) > 11.5$], in order to stack galaxies with similar properties in the $z - \log (M_*/\mathrm{M_\odot})$ plane. We shifted the individual spectra to their rest-frame, and we normalised their flux within $5150-5500\,\text{\AA}$.  The chosen stacking redshift bin is wide enough to include spectra with slightly different observed wavelength ranges, allowing for optimal removal of bad pixels and sky-subtraction residuals. For each stacked spectrum, we generated $201$ bootstrap realisations by randomly resampling, with replacement, the set of 50 spectra in each bin. In each realisation, 50 spectra were drawn, allowing individual spectra to appear multiple times, and co-added following the same procedure adopted for the original stack. The dispersion among the resulting stacks was then used to estimate robust pixel–by–pixel uncertainties.
From this step we obtained $2835$ stacked spectra in the mass bin $11.3< \log (M_*/\mathrm{M_\odot}) < 11.5$ and $842$ in the mass bin $\log (M_*/\mathrm{M_\odot}) > 11.5$.

\begin{figure}
   \centering
   \includegraphics[width=0.5\textwidth]{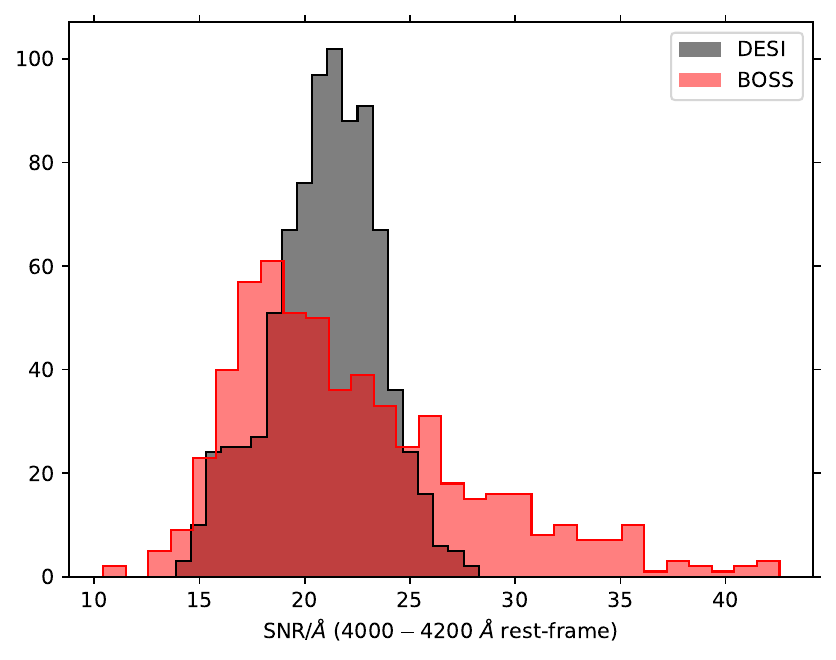}
      \caption{SNR distributions for the DESI stacked spectra, in black, and BOSS, in red, in the rest-frame range $4000-4200\,\text{\AA}$.
              }
         \label{fig:snrdesiboss}
         
   \end{figure}

Regarding the BOSS sample, the median SNR of individual spectra is around $8/\text{\AA}$ in the rest-frame range $4000-4200\,\text{\AA}$, therefore we stacked groups of $10$ spectra in order to achieve a SNR comparable to that of the DESI stacked spectra. From this step we obtained $581$ BOSS stacked spectra. Figure~\ref{fig:snrdesiboss} shows the SNR distributions for the DESI and BOSS stacked spectra in the rest-frame range $4000-4200\,\text{\AA}$. The BOSS distribution shows a more extended high S/N tail than DESI, indicating a larger fraction of spectra at high S/N, while the median values are consistent to SNR $\sim 21.5/\text{\AA}$.
The choice of the number of spectra to be stacked ensures reliable measurements of the stellar population properties and their intrinsic spread across the redshift range by stacking only the minimum number of spectra required. This approach preserves sensitivity to the intrinsic variation within the sample, which would otherwise be lost if all available spectra were combined into a single very high-SNR stack.

\section{Analysis}
Our goal is to characterise the star formation and chemical enrichment history of the most massive quiescent galaxies through their light-weighted age, stellar metallicity, and [$\alpha$/Fe] abundance. 
In this section, we describe in detail the steps of the analysis performed on our sample. Specifically, 
in Sect.~\ref{subsec:models} we introduce the two stellar population models used to analyse the selected galaxies. In Sect.~\ref{subsec:smiles_analysis} and Sect.~\ref{subsec:classicindex} we describe in detail how we obtained the estimate of stellar kinematics, the extraction of the stellar components and the measurements of the stellar population parameters using Monte Carlo techniques. Finally we present how we assess the stellar population parameters as function of redshift using the hierarchical bayesian modeling (Sect.~\ref{subsec:hierarc}).

\label{sec:analysis}
\subsection{Stellar population models}
\label{subsec:models}
We used two spectrophotometric models to measure the light-weighted age, stellar metallicity and [$\alpha$/Fe] from the stacked spectra. The first set corresponds to the semi-empirical Simple Stellar Populations (SSPs) models from \cite{knowles2023} (sMILES hereafter). 
These SSP models adopt the BaSTI evolutionary tracks \citep{pietrinferni2004large,pietrinferni2006large} and the MILES empirical stellar library \citep[$3540.5 < \lambda < 7350.2\,\text{\AA}$;][$2.5\,\text{\AA}$ FWHM resolution]{sanchez2006medium,falcon2011updated}. We assumed a Kroupa initial mass function \citep[IMF,][]{kroupa2001}. The synthetic library provides $2650$ SSPs unevenly spaced in linear age and [M/H], covering $53$ ages from $0.03$ Gyr to $14$ Gyr, $10$ metallicities from [M/H]$\ = -1.79$ to [M/H]$\ = 0.26$ dex, being the solar abundance Z$_\odot = 0.0198$, and five [$\alpha$/Fe] abundances from $-0.2$ to $0.6$ dex. 
The second set is that of \cite{thomas2011} (hereafter TMJ).
These models are based on the Cassisi evolutionary tracks \citep{cassisi1997}, the MILES empirical stellar library, and assume a Kroupa initial mass function. For these models, the synthetic library provides predictions of $25$ Lick indices  for each SSP \citep{worthey1994old}. The SSPs are unevenly spaced in linear age and [M/H], covering $20$ ages from $0.1$ Gyr to $15$ Gyr, $6$ metallicities from [M/H]$\ = -2.25$ to [M/H]$\ = 0.67$ dex (with solar abundance Z$_\odot = 0.02$), and five [$\alpha$/Fe] abundance ratio from $-0.3$ to $0.5$ dex.

In the following, we use these two stellar population models to perform parallel analyses of our sample, allowing us to compare the derived stellar population parameters under different modelling assumptions. We present the details of the methodology and results obtained with each models in the subsequent sections.

\subsection{sMILES analysis}
\label{subsec:smiles_analysis}
\subsubsection{Kinematics and residual emission lines subtraction}
\label{subsubsection:kinematicsmiles}
For each stacked spectrum in the sample, we derived the kinematic parameters (i.e., recession velocity residuals and velocity dispersion) using the latest version of the pPXF code\footnote{https://pypi.org/project/ppxf/}\citep{cappellari2004parametric,cappellari2023}. We adopted stellar templates from the MILES stellar spectral library and performed the fit over the rest-frame wavelength range $3800-5500\,\text{\AA}$. The synthetic templates were shifted to match each galaxy’s residual recession velocity and convolved with the corresponding stellar velocity dispersion. We masked the Balmer line regions to avoid that possible small residuals of emission lines affect the fit of the continuum.
Although we applied the spectroscopic criteria outlined in Sect.~\ref{sec:datasel} to select our sample galaxies, since the individual spectra have low SNR, they might be affected by weak nebular emission filling in the cores of absorption features. To mitigate this effect, we analysed the stacked spectra for the presence of residual nebular emission in the Balmer lines. Thanks to their high S/N ratio, even small emission-line contributions can be reliably identified and assessed.
This is especially important for the $H\beta$ region, where even modest emission can bias the inferred stellar ages. 
We therefore performed an additional two-step pPXF spectral fitting to extract the nebular emission component from each stacked spectrum, fixing the kinematics to the results obtained in the first stage. 
In the first step, we fitted only H$\beta$ emission line within the wavelength range $4000\,\text{\AA} < \lambda < 5500\,\text{\AA}$, to obtain a reliable estimate of the nebular kinematics. In the second step, using the same wavelength range, we simultaneously fitted $H\delta$, $H\gamma$ and $H\beta$ emission lines while keeping the nebular kinematics fixed to the values obtained in the first step. 
We retained only those fits for which the recovered Balmer fluxes follow the expected sequence $H\delta < H\gamma < H\beta$. If this condition was not met, we fixed the Balmer emission-line ratios to the case-B recombination values linked to the measured H$\beta$ flux \citep{osterbrock1989}. Finally, we subtracted the best-fit nebular component from the global spectrum, yielding a clean stellar component spectrum. We then performed $1000$ realisations of the nebular best-fit by perturbing the emission line fluxes according to their associated flux uncertainties, in order to estimate the pixel-wise errors in the $H\delta$, $H\gamma$, and $H\beta$ regions of the nebular component. These estimated uncertainties were then added in quadrature to those of the stacked spectrum.
\subsubsection{Full-index fitting}
\label{subsec:fullindex}

\begin{figure*}
   \centering
   \includegraphics[width=0.95\textwidth]{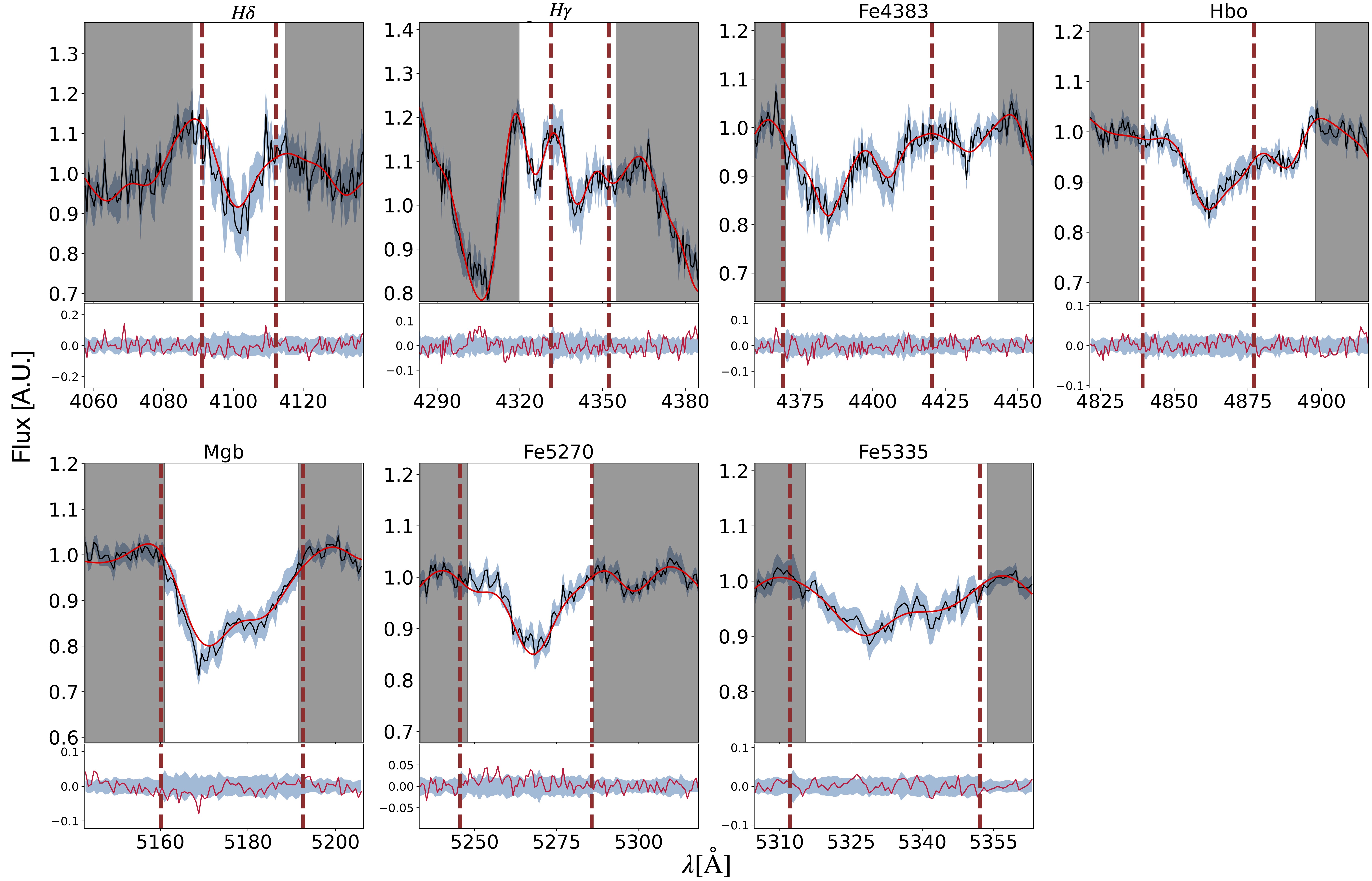}
      \caption{FIF application on the selected indices for a stacked spectrum in the DESI sample. The vertical dashed brown lines indicate the feature boundaries for each index while the grey shaded area represent the pseudo-continua regions used for normalisation. In the upper subplots, the black lines and blue shaded regions correspond to the stacked spectrum and its associated uncertainty, respectively. The solid red line represents the best-fit derived from the posterior distribution. The red lines in the lower subplots show the residuals between the observed spectrum and the best fit, with the blue shaded region indicating the relative uncertainties of the stacked spectrum.
              }
         \label{fig:confr_fif}
         
   \end{figure*}

We adopted the full-index fitting approach \citep[FIF,][]{martin2015stellar,ditrani2025} to compare the observed spectra with the synthetic spectral templates. Unlike the more classic index fitting approach, FIF compares the flux within a specific absorption feature (with respect to its continuum value) pixel by pixel rather than averaging it. This pixel-level comparison within the index window is more effective in breaking the degeneracy between age and metallicity compared to the classical index analysis, as it accounts not only for the strength of the absorption feature but also for its detailed shape, which provides additional information about the stellar population parameters. \citep[][]{martin2019fornax,ditrani24}. 
We applied the FIF method to a comprehensive set of spectral indices sensitive to the age, stellar metallicity and [$\alpha$/Fe], listed in Table~\ref{table:indices}. Because our fitting scheme includes Balmer lines together with Fe– and Mg–sensitive indices, the inferred [$\alpha$/Fe] ratio effectively traces the [Mg/Fe] abundance ratio. Magnesium is the only $\alpha$–element directly constrained by our index set, whereas the remaining $\alpha$–elements (e.g. O, Si, Ca, Ti) have negligible effects on the indices used. As a consequence, the [$\alpha$/Fe] values returned by the models should be interpreted as [Mg/Fe] proxy rather than as fully general $\alpha$–enhancement measurements. Moreover, since we are fitting spectra with a single SSP, the derived age corresponds to the SSP-equivalent age, i.e. the age of the SSP that best reproduces the observed spectrum. In practice, this quantity closely traces the light-weighted age, since the fit is driven by the luminosity-weighted contribution of the underlying stellar populations.
Figure~\ref{fig:confr_fif} shows an example of the FIF approach applied to one of the DESI stacked spectra. The best-fit template closely matches each spectral feature, capturing detailed information from both their depth and shape.

\begin{table*}
\caption{Spectral indices used for our analysis.}
\centering  
\begingroup
\begin{tabular}{lllllll}
\hline\hline      
Index & Blue pseudocontinuum & Central feature & Red pseudocontinuum & Fit$_\text{sMILES}$ & Fit$_\text{TMJ}$ & Ref. \\
\hline
H$\delta_F$ & 4057.25-4088.50 & 4091.00-4112.25 & 4114.75-4137.25 & [$\alpha$/Fe] = 0 & variable [$\alpha$/Fe] & 1 \\
H$\gamma_F$ & 4283.50-4319.75 & 4331.25-4352.25 & 4354.75-4384.75 & [$\alpha$/Fe] = 0 & variable [$\alpha$/Fe] &  1 \\
Fe4383 & 4359.125-4370.375 & 4369.125-4420.375 & 4442.875-4455.375 & variable [$\alpha$/Fe] & variable [$\alpha$/Fe] & 1 \\
$H\beta_o$ & 4821.175-4838.404 & 4839.275-4877.097 & 4897.445-4915.845 & [$\alpha$/Fe] = 0 & variable [$\alpha$/Fe] & 2 \\
Mgb & 5142.625-5161.375 & 5160.125-5192.625 & 5191.375-5206.375 & variable [$\alpha$/Fe] & variable [$\alpha$/Fe] &1  \\
Fe5270 & 5233.150-5248.150 & 5245.650-5285.650 & 5285.650-5318.150 & variable [$\alpha$/Fe] & variable [$\alpha$/Fe] &1 \\
Fe5335 & 5304.625-5315.875 & 5312.125-5352.125 & 5353.375-5363.375  & variable [$\alpha$/Fe] & variable [$\alpha$/Fe] & 1 \\
\hline

\end{tabular}
\vspace{2mm}
\parbox{0.9\linewidth}{\small  (1)\cite{worthey1994old}; (2) \cite{cervantes2009optimized}.}

\endgroup
\label{table:indices}
\end{table*}

\begin{figure}
   \centering
   \includegraphics[width=0.5\textwidth]{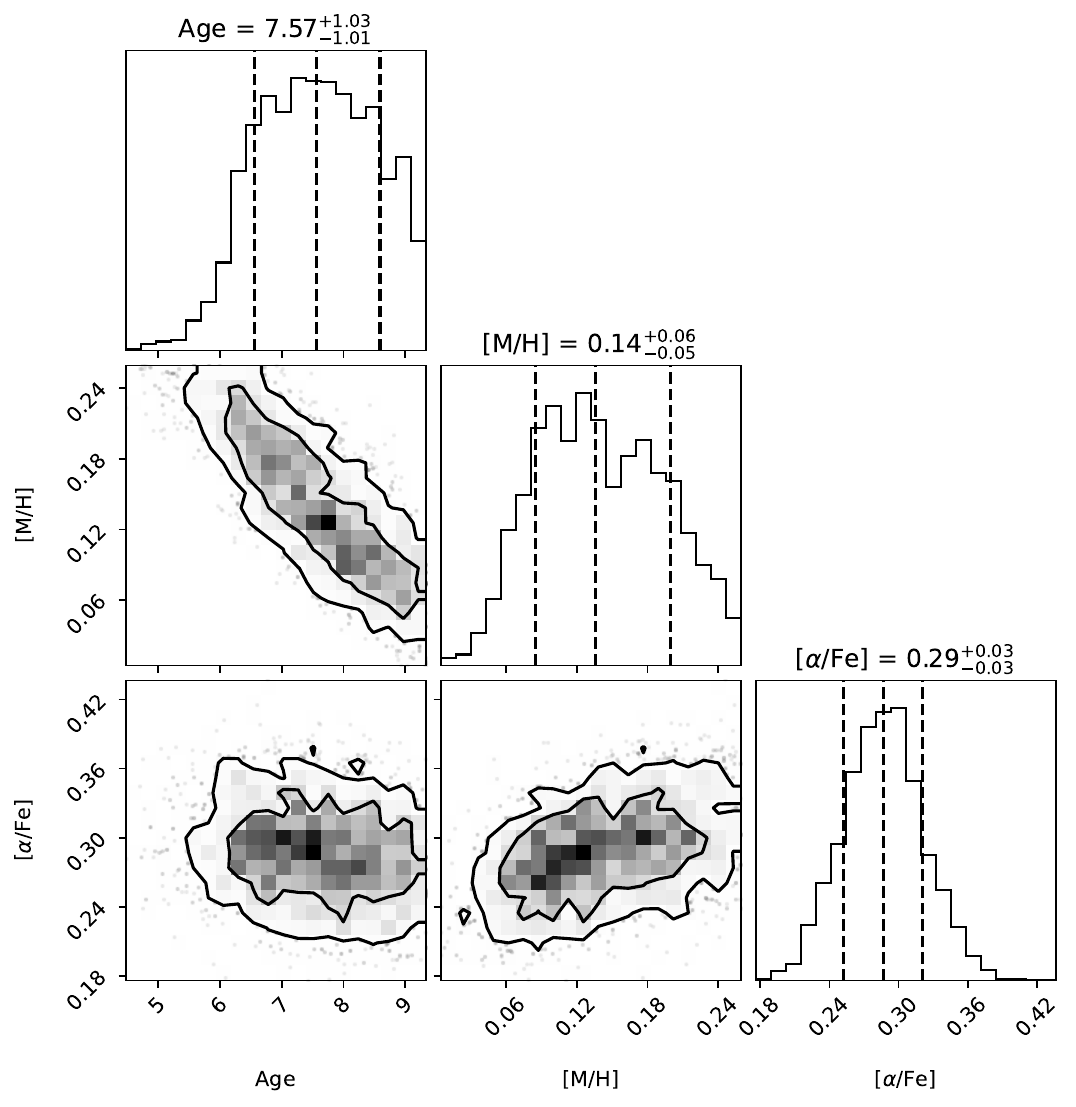}
      \caption{Example of the joint and marginal posterior distributions of light-weighted age, [M/H], and [$\alpha$/Fe] for a DESI stacked spectrum at $z = 0.4$ and $\log (M_*/\mathrm{M_\odot}) = 11.6$. Contours represent the $68\%$ and $95\%$ probability level. The $16\%$, $50\%$ and $84\%$ intervals are indicated as dashed lines.}
         \label{fig:corner_all}
   \end{figure}

Modelling the effect of [$\alpha$/Fe] $ \neq 0$ \citep[see][]{vazdekis2015evolutionary,knowles2023} models is known to be problematic, as it relies on stellar models computed under the assumption of local thermodynamic equilibrium (LTE), which may underestimate Balmer line depths in $\alpha$-enhanced or hot stellar populations, where non-LTE effects are known to be non-negligible. Moreover, in practice, the effect of varying the $\alpha$-elements might be balanced by variations of other elements (e.g. C).
To mitigate these effects, for sMILES we fitted the set of indices [H$\delta_F$, H$\gamma_F$, $H\beta_o$] adopting SSP templates corresponding to [$\alpha$/Fe] $= 0$, while we fitted the remaining indices using SSPs with variable [$\alpha$/Fe] abundance ratio.
We computed the total posterior probability distribution using the likelihood given by $\mathcal{L} = e^{-\chi^2/2}$, with
\begin{equation}
    \chi^2 = \sum_{i}{\left(\frac{F_{obs_i}-F_{syn_i}}{\sigma_{obs_i}}\right)}^2
\end{equation}
where F$_{syn_i}$ is the flux of the synthetic spectrum along the feature of each index, and F$_{obs_i}$ is the flux of the observed spectrum with the error $\sigma_{obs_i}$. 
We derived posterior probability distributions and the Bayesian evidence using the nested sampling Monte Carlo algorithm
MLFriends \citep{2016S&C....26..383B,2019PASP..131j8005B} using the UltraNest\footnote{\url{https://johannesbuchner.github.io/UltraNest/}} package \citep{2021JOSS....6.3001B}. We assumed uniform prior for all the parameters considered, summarised in Table~\ref{table:params}. 
Figure~\ref{fig:corner_all} shows an example of the joint and marginal probability distributions for all the fitted parameters.
\begin{table}
\caption{The $3$ free parameters fitted to our spectroscopic data, along with their associated prior distributions.}
\centering  
\begingroup
\begin{tabular}{lllll}
\hline\hline      
Parameter & Unit & Range & Prior & model \\
\hline
 Age$_{\text{model}}$ & Gyr & (0, Age$_{\text{Universe}}$) & Uniform & sMILES  \\
 $\text{[M/H]}$ & dex & (-1.79, 0.26) & Uniform & sMILES \\
$\text{[$\alpha$/Fe]}$ & dex & (-0.2, 0.6) & Uniform & sMILES  \\
\hline
 Age$_{\text{model}}$ & Gyr & (0, Age$_{\text{Universe}}$) & Uniform & TMJ  \\
 $\text{[M/H]}$ & dex & (-2.25, 0.67) & Uniform & TMJ \\
$\text{[$\alpha$/Fe]}$ & dex & (-0.3, 0.5) & Uniform & TMJ  \\
\hline
\end{tabular}
\endgroup
\label{table:params}
\end{table}

This technique of full-index fitting combined with the nested sampling algorithm has been implemented in a python routine, named \textbf{INFUSE}\footnote{\url{https://infuse-fit.readthedocs.io/en/latest/}} (Full-INdex Fitting for Uncovering Stellar Evolution), which we publicly release with this work. 

\subsection{TMJ analysis}
\label{subsec:classicindex}
The analysis based on the TMJ models follows the same general procedure adopted for the sMILES analysis, with the important distinction that TMJ provides predictions for absorption-line indices rather than synthetic spectra. Since the spectral resolution of TMJ cannot be adapted to the effective resolution of our stacked spectra, the latter must be degraded to the model resolution to ensure consistency with the model predictions.
For this reason, before performing any kinematic measurement and emission-line correction, we first homogenised the stacked spectra by convolving them to the fixed spectral resolution of the TMJ models (FWHM = $2.51\,\text{\AA}$).
After that, we carried out the same procedure adopted for the sMILES analysis to measure the stellar kinematics and perform the removal of emission line residuals (see Sect.~\ref{subsubsection:kinematicsmiles}). 

\begin{figure*}
   \centering
   \includegraphics[width=0.95\textwidth]{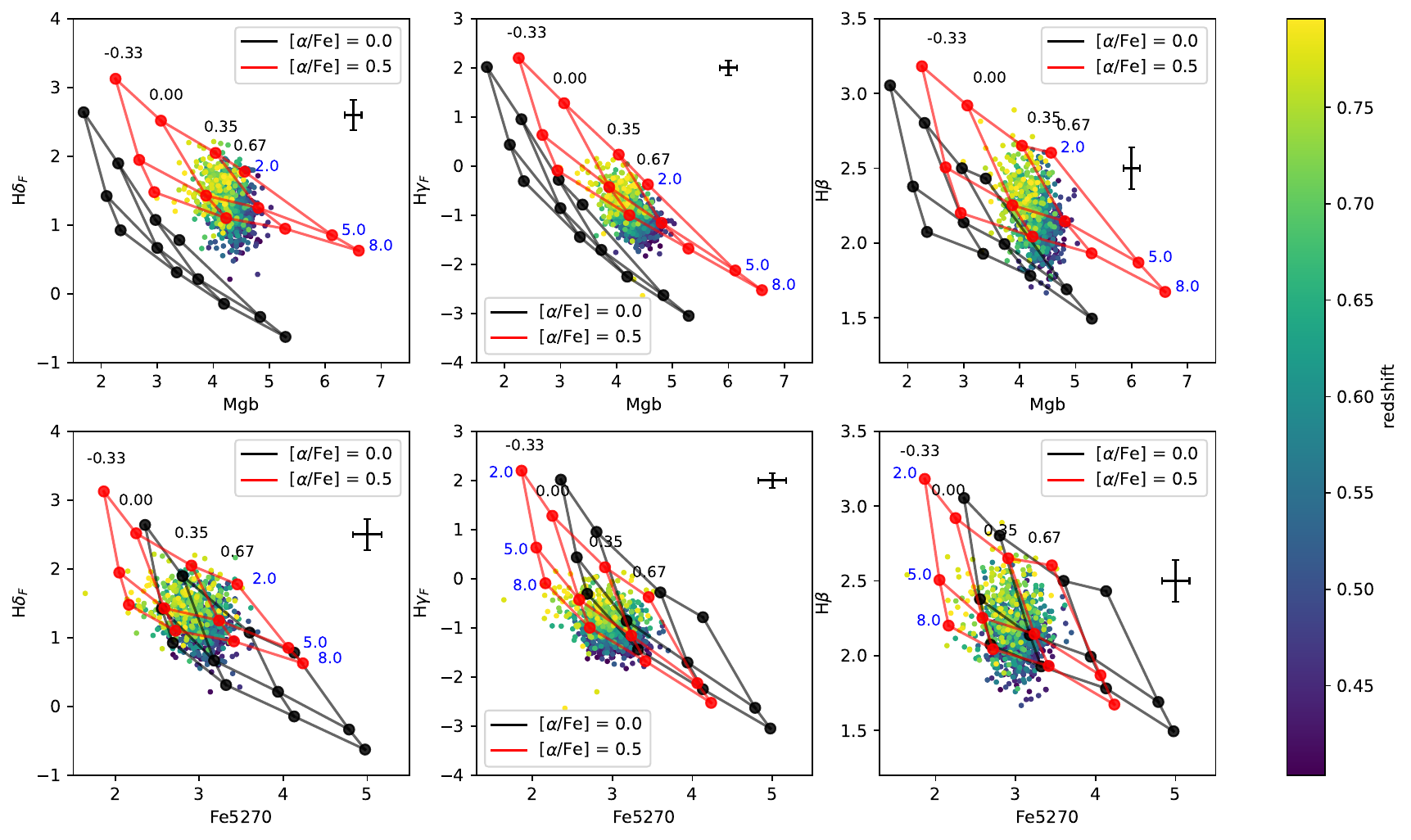}
      \caption{Balmer line indices H$\delta_F$, H$\gamma_F$, and H$\beta$ as functions of Mgb (top panels) and Fe$5270$ bottom panels. Red and black lines are the $\alpha$/Fe-enhanced and the solar-scales values for the TMJ models, respectively. Models for the ages 2, 5, and 8 Gyr, and the metallicities [M/H] $= -0.33$, $0.0$, $0.35$, $0.67$ dex are shown.
      The circles marks the measured indices on the DESI stacked spectra, colour-coded as function of the redshift. The error bars denote average $1\sigma$ errors.
              }
         \label{fig:indexindex_tmj}
         
   \end{figure*}

Since the TMJ models provide predictions for absorption-line indices, we carried out the stellar population analysis using the classical index–strength approach. For consistency with the sMILES analysis, we measured the same set of indices listed in Table~\ref{table:indices} on each stacked spectrum. In this analysis we considered the effect of [$\alpha$/Fe] also on higher-order Balmer lines (see~\ref{table:indices}), for which the TMJ models provide a robust calibration \citep[][]{thomas2004}.
Because the observed spectra are broadened by the stellar velocity dispersion, we corrected the measured indices for this effect following the procedure described in \cite{thomas2010} \citep[see also][]{kuntschner2004}. We derived the correction factor using the best-fitting pPXF template obtained for each stacked spectrum, as follows: first,
we measure each index on the best-fitting template spectrum degraded only to the TMJ spectral resolution (FWHM = $2.51\,\text{\AA}$) and with zero velocity dispersion; second, convolve the same best-fitting template with the stellar velocity dispersion obtained from pPXF fit, and measure the indices again. The ratio:
\begin{equation}
    C = \frac{I_\text{bestfit}}{I_{\text{bestfit, broadened}}}
\end{equation}
provides the multiplicative correction factor applied to the observed indices. This procedure ensures that the observed line strengths are brought onto the TMJ index system at fixed resolution and zero velocity dispersion, allowing a direct comparison with the model predictions. Figure~\ref{fig:indexindex_tmj} shows the Balmer line indices H$\delta_F$, H$\gamma_F$, and H$\beta$ as functions of Mgb and Fe$5270$ for TMJ models, together with the corresponding indices measured on the DESI stacked spectra as a function of the redshift. The comparison clearly illustrates that the models span the region occupied by the data, and highlights the important role that abundance ratios might have in interpreting these spectral indices. 

We computed the total posterior probability distribution using the likelihood given by $\mathcal{L} = e^{-\chi^2/2}$, with: 
\begin{equation}
    \chi^2 = \sum_{i}{\left(\frac{I_{obs_i}-I_{syn_i}}{\sigma_{obs_i}}\right)}^2
\end{equation}
where $I_{obs_i}$ is the \textit{i}th spectral index measured in the stacked spectrum, $I_{syn_i}$ is the index prediction of the models, and $\sigma_{obs_i}$ is the observational error of the \textit{i}th spectral index. The observational errors have been evaluated as the standard deviation of the distribution of $1000$ random Gaussian realisations of the perturbed stacked spectrum. We derived posterior probability distributions  using the UltraNest algorithm, assuming uniform prior for all the parameters considered, summarised in Table~\ref{table:params}.
\subsection{Hierarchical Bayesian Modeling}
\label{subsec:hierarc}
\begin{figure}
   \centering
   \includegraphics[width=0.5\textwidth]{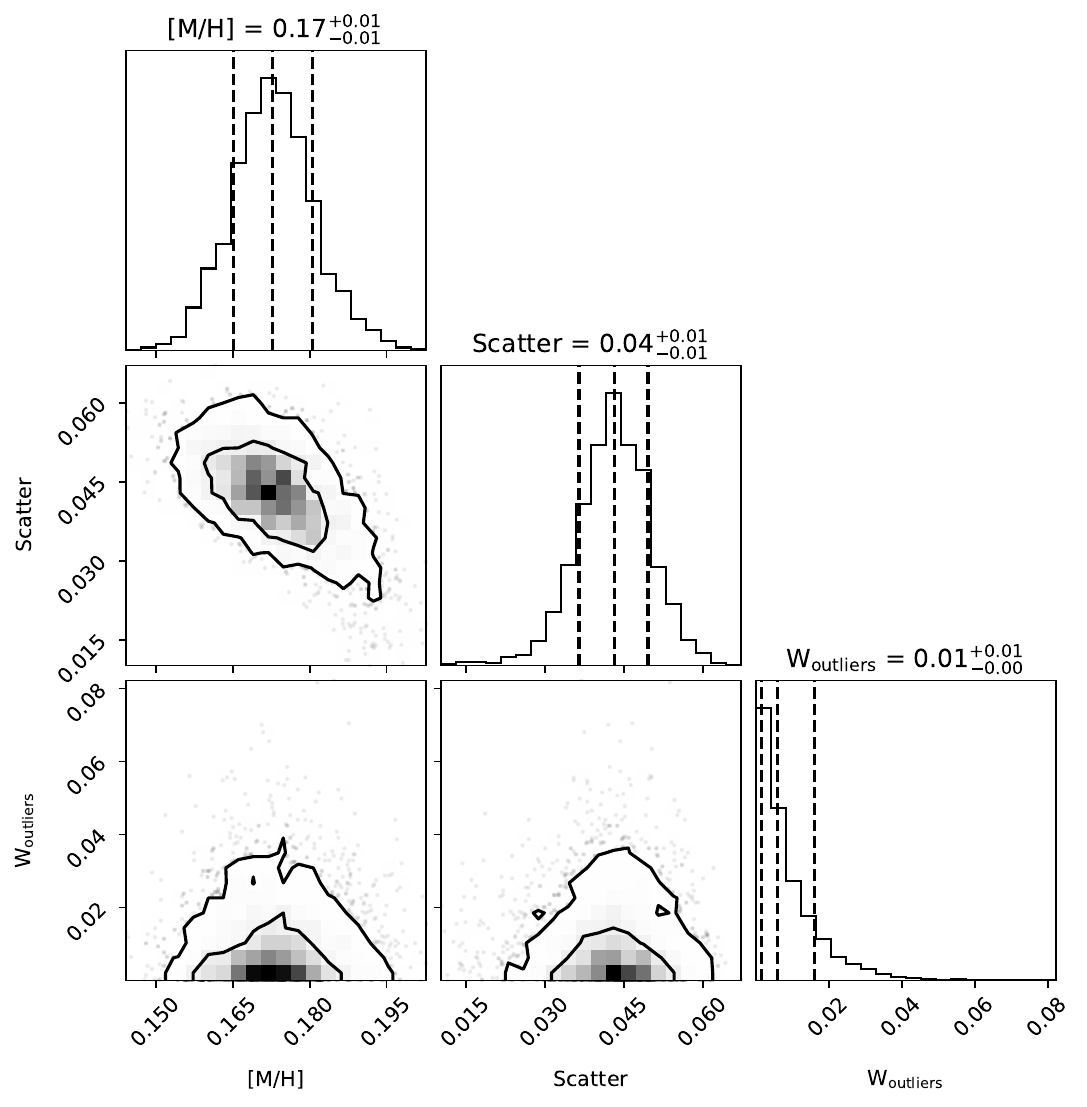}
      \caption{Corner plot summary of the posterior distributions of the [M/H], its scatter, and the fraction of outliers obtained for all the stacked spectra contained in the redshift range $0.45 < z < 0.5$ and [$\log (M_*/\mathrm{M_\odot}) > 11.5$] bin. Contours are at the $68\%$ and $95\%$ probability. The $16\%$, $50\%$ and $84\%$ intervals are indicated by dashed lines.}
         \label{fig:corner_hierarc}
   \end{figure}

To assess the stellar population parameters as a function of redshift, we grouped the stacked spectra into redshift bins. We adopted bins of width $\Delta z = 0.05$, and for each bin we used the corresponding stacked spectra falling within that redshift interval, to ensure to have a statistically significant sample in each bin. We then combined the posteriors of each stacked spectrum using the hierarchical bayesian modeling. In the hierarchical framework, our models consist of two levels: the first level involves the measurements of the parameters for each individual stacked spectrum, while the second level describes how the measurements are distributed within each redshift bin. Differently from the classic stacked spectra fit, this approach has several advantages as it avoids introducing correlated noise caused by smoothing the individual spectra to a common velocity dispersion and by the continuum interpolation with polynomials \citep[see Appendix B in][]{beverage2023}.
Following the 'a posteriori' approach \citep[e.g.][]{beverage2023,ditrani2025}, as first level of the models we computed the posterior probability distribution of each physical parameter listed in Table~\ref{table:params} for each individual stacked spectrum as detailed in Sect.~\ref{subsec:fullindex} and Sect.~\ref{subsec:classicindex}. Then, as a second level of modelling, we fitted the distribution of the posteriors of each parameter with a Gaussian function plus an outlier model in each redshift bin. 
This approach provides a mean value for each parameter in each selected redshift bin, the intrinsic scatter of our sample, along with a reliable estimate of the uncertainties. We assumed the same prior listed in Table~\ref{table:params} for each parameter, then we applied a logarithmically uniform prior for the intrinsic scatter between $0.01$ and $10$ dex, and a uniform prior for the fraction of outliers within $0-20\%$. Figure~\ref{fig:corner_hierarc} shows an example of the posterior probability distribution of the [M/H] parameter, its scatter, and the fraction of outliers for the DESI spectra in the redshift bin $0.45 < z < 0.5$ and mass bin [$\log (M_*/\mathrm{M_\odot}) > 11.5$]).

\section{Results}


In the following we present the results obtained using both the sMILES and TMJ models. Specifically, we discuss the stellar population parameters of the DESI stacked spectra in two distinct mass bins, and the redshift evolution of these parameters of the most massive galaxies across the full BOSS and DESI redshift range.

When comparing stellar population parameters derived using the TMJ and sMILES models, one should be aware of an important difference. Both frameworks provide a total metallicity [M/H], but the resulting values are not directly comparable, as each assumes a different $\alpha$-elements mixture at fixed iron abundance. 
In practice, while Fe and Mg are the only elements directly and robustly constrained by the spectral indices used in this work, the remaining metal content is model dependent and inferred through the adopted $\alpha$-element pattern.
Indeed, in the sMILES models the [$\alpha$/Fe] ratio accounts for variations in the $\alpha$–elements O, Ne, S, Mg, Si, Ca, and Ti. In contrast, in the TMJ models the definition of the $\alpha$–enhanced mixture additionally includes C, N, and Na, whose abundances are modified alongside the other $\alpha$–elements. The relation between [Fe/H], [$\alpha$/Fe] and [M/H] differs substantially between the two sets of models:
\begin{align}
\hspace{0pt}\left[ \mathrm{M/H} \right]_{\text{sMILES}} &= \left[  \mathrm{Fe/H} \right] + 0.66\left[ \mathrm{\alpha/Fe} \right] + 0.20\left[ \mathrm{\alpha/Fe} \right]^2 \\
\hspace{0pt}\left[ \mathrm{M/H} \right]_{\text{TMJ}} &= \left[ \mathrm{Fe/H} \right] + 0.94\left[ \mathrm{\alpha/Fe} \right]
\end{align}
As a consequence, the [M/H] parameter directly returned by the fit cannot be compared at face value. In contrast, [Fe/H] is the most robust quantity to compare between the two models, because it is directly constrained by Fe-sensitive spectral lines and is only weakly affected by model-dependent assumptions about $\alpha$-element enhancement. 
For this reason, in the following we compare light-weighted age, and separately [Fe/H], and [$\alpha$/Fe] between TMJ and sMILES, since [M/H] is a different combination of the two quantities above for the two models.

\begin{figure*}
   \centering
   \includegraphics[width=0.9\textwidth,height=0.8\textheight,keepaspectratio]{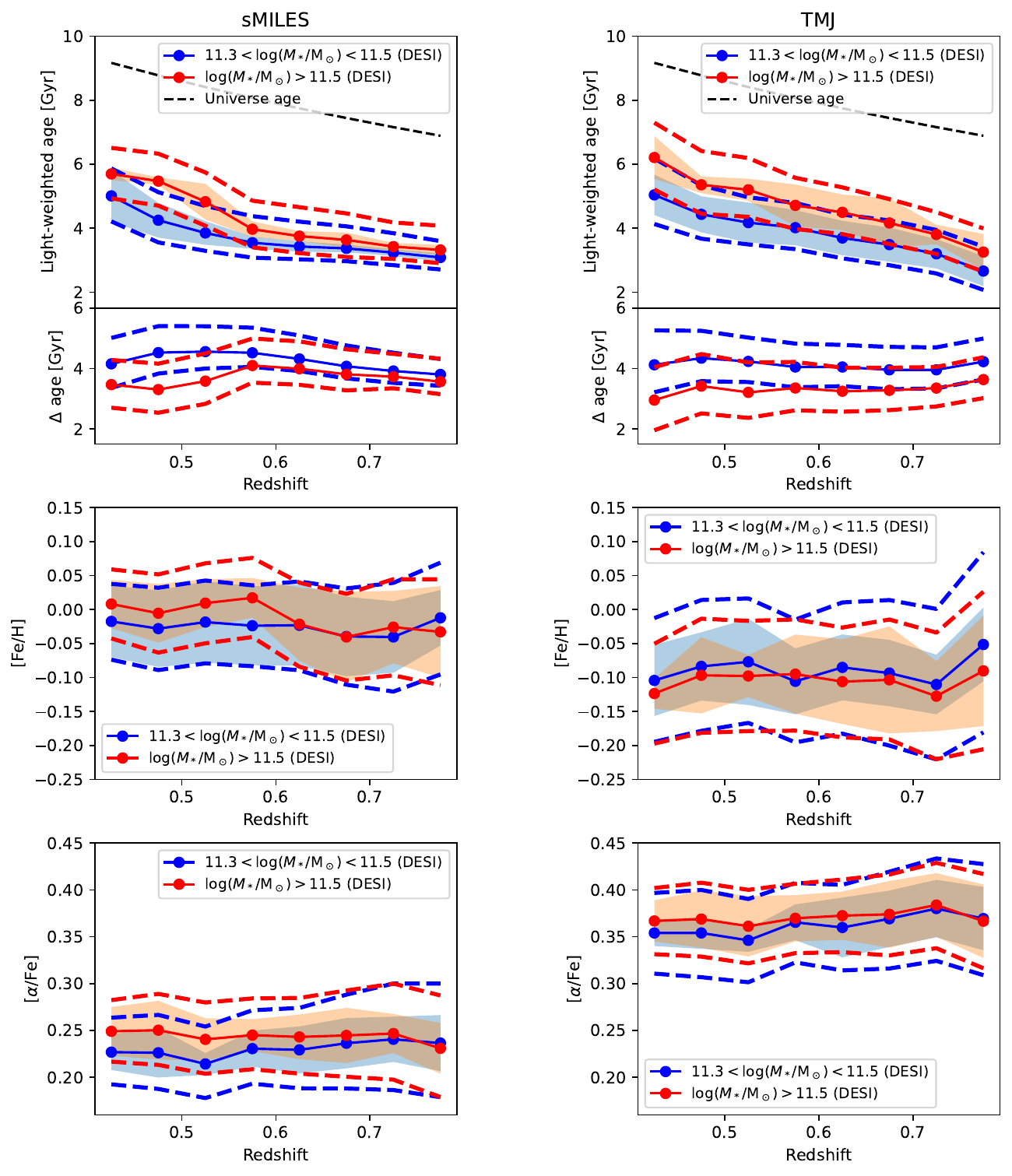}
      \caption{Evolution of the light-weighted age, [Fe/H] and [$\alpha$/Fe] in the redshift range $0.4<z<0.8$ in the mass bins [$11.3 < \log (M_*/\mathrm{M_\odot}) < 11.5$, $\log (M_*/\mathrm{M_\odot}) > 11.5$]. Each top panel includes a sub-panel showing the difference between the age of the Universe and the light-weighted age.
      The three panels on the left show the results obtained using sMILES, while the three panels on the right show the corresponding results from TMJ.
      The data points represent the median value in each redshift bin, the dashed lines cover the typical uncertainties of the stacked spectra, and the shaded regions correspond to the intrinsic scatter within each bin. The dashed black line in the top panel represents the age of the Universe at each redshift. }
         \label{fig:resultsdesitwomass}
         
   \end{figure*}

Figure~\ref{fig:resultsdesitwomass} shows the evolution of the light-weighted age, [Fe/H] and [$\alpha$/Fe] in the redshift range $0.4<z<0.8$ for the DESI sample in the two mass bins [$11.3 < \log (M_*/\mathrm{M_\odot}) < 11.5$, and $\log (M_*/\mathrm{M_\odot}) > 11.5$]. As described in Sect.~\ref{subsec:hierarc}, we estimated the mean and intrinsic scatter of each parameter in redshift bins of width $\Delta z = 0.05$. The distributions of the stellar population parameters follow consistent trends across the entire redshift range, for the two mass bins, suggesting similar star formation histories. These trends are also consistent across the sMILES and TMJ models, with the most significant differences found in the light-weighted age and [$\alpha$/Fe] values. Within each redshift bin, we found no evidence for significant intrinsic scatter beyond the observational uncertainties of the individual stacked spectra, suggesting that the galaxy population is nearly homogeneous.
Overall, galaxies in the higher mass bin exhibit systematically older ages and higher [$\alpha$/Fe] across the entire redshift range covered by DESI, while [Fe/H] is consistent in the two mass bins, although these differences remain within the $1\sigma$ uncertainties. 

\begin{figure*}
   \centering
   \includegraphics[width=0.9\textwidth,height=0.8\textheight,keepaspectratio]{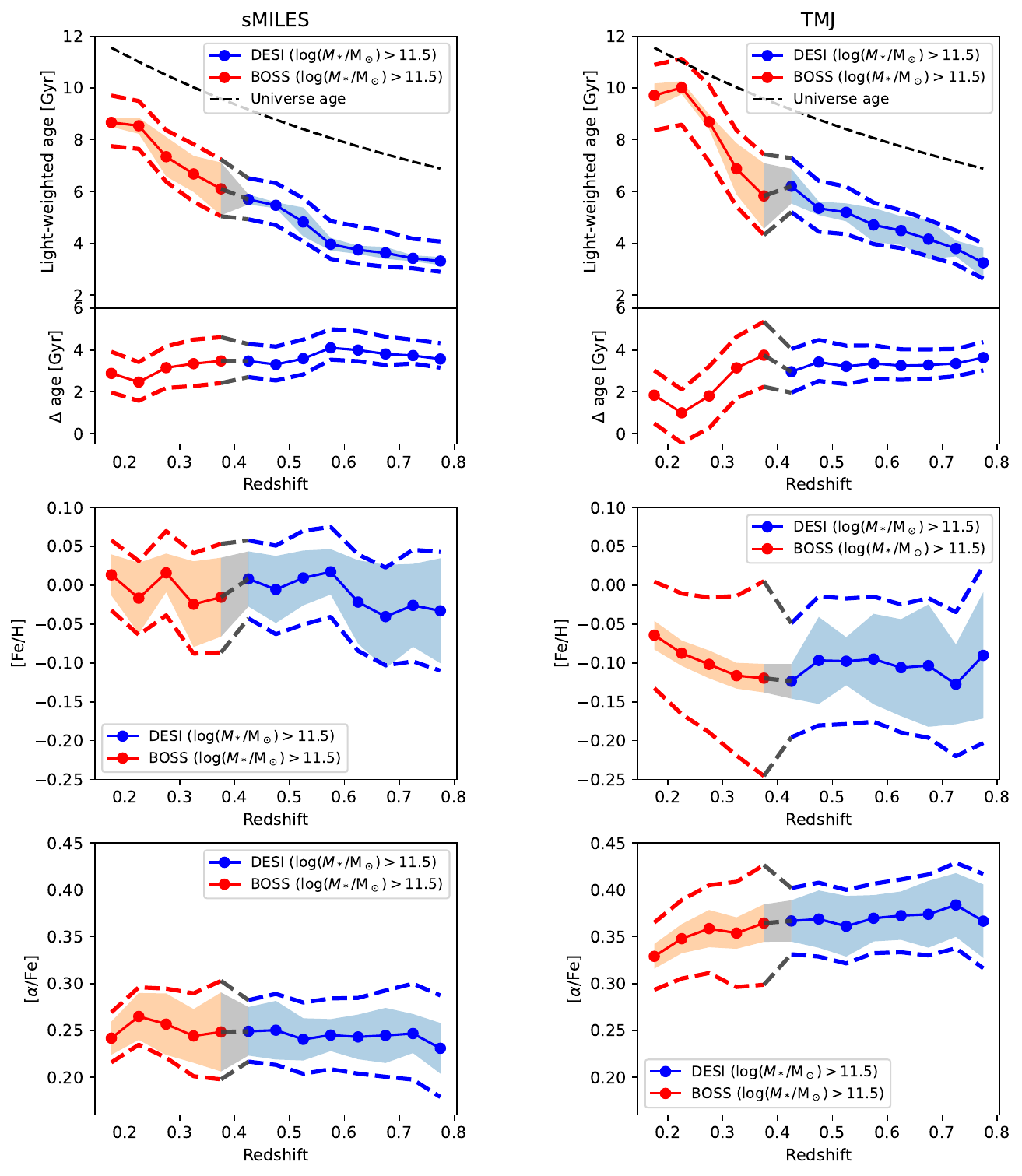}
      \caption{Evolution of the light-weighted age, [Fe/H] and [$\alpha$/Fe] in the redshift range $0.15<z<0.8$. Each top panel includes a sub-panel showing the difference between the age of the Universe and the light-weighted age. The three panels on the left show the results obtained using sMILES, while the three panels on the right show the corresponding results from TMJ. The data points represent the median value in each redshift bin, the dashed lines cover the typical uncertainties of the stacked spectra, and the shaded regions correspond to the intrinsic scatter within each bin. The dashed black line in the top panel represents the age of the Universe at each redshift.
              }
         \label{fig:resultsbossdesi}
         
   \end{figure*}

Figure~\ref{fig:resultsbossdesi} shows the evolution of the light-weighted age, [Fe/H] and [$\alpha$/Fe] in the redshift range $0.15<z<0.8$ of stacked spectra of galaxies of the BOSS and DESI samples, for the common mass bin [$\log (M_*/\mathrm{M_\odot}) > 11.5$]. Also in this case, we estimated the mean and intrinsic scatter of each parameter in redshift bins of width $\Delta z = 0.05$. Overall, the two surveys show consistent evolutionary trends and all parameters exhibit clear correlations with redshift. Despite being independent datasets, the BOSS and DESI results agree well in the redshift range where they overlap, supporting a consistent selection function of LRGs and similar mass estimates in both surveys (as compared in Sect.~\ref{sec:datasel}). For both the BOSS and DESI results we found no evidence for significant intrinsic scatter beyond the observational uncertainties of the individual stacked spectra. In both samples, galaxies exhibit an evolution of the light-weighted age consistent with a pure passive evolution, with a nearly constant offset between the age of the Universe and the measured stellar ages. The galaxies' [Fe/H] and [$\alpha$/Fe] show an overall flat trend towards lower redshift. 
These trends are recovered using both the sMILES and TMJ models, even if the two analyses yield different absolute values for [Fe/H] and [$\alpha$/Fe]. In particular, the sMILES models return systematically higher [Fe/H] and lower [$\alpha$/Fe] compared to TMJ, while preserving the same redshift dependence.


\label{sec:result}

\section{Most massive quiescent galaxies in the I\lowercase{llustris}TNG simulation}
\label{sec:tngcomparison}
The IllustrisTNG simulations \citep[][TNG hereafter]{springel2018,pillepich2018,nelson2019} are a series of cosmological gravity+magnetohydrodynamical simulations that model a range of physical processes for the formation of galaxies. Each TNG simulation includes a comprehensive model of galaxy formation and evolution from $z = 127$ to $z = 0$, generating several snapshots across the cosmic time. The initial conditions of the TNG snapshots have been initialised at $z = 127$ assuming a matter density $\Omega_m = \Omega_{dm} + \Omega_b = 0.3089$, baryonic density $\Omega_b = 0.0486$, cosmological constant $\Omega_\Lambda = 0.6911$, Hubble constant $H_0 = 100 h$ km s$^{-1}$ Mpc$^{-1}$ with $h = 0.6774$ \citep{planck2016}.
In this work, we considered the largest TNG box: TNG300-1, which has a side length of $300$ Mpc.
For our analysis, we considered snapshots within the redshift range $0.15 < z < 0.8$, in order to compare the stellar population parameters of simulated galaxies with those obtained from the spectral analysis of our sample. Specifically, we used the full output snapshots 
 at $z = [0.7, 0.5, 0.4, 0.3, 0.2]$, as the [$\alpha$/Fe] information is only available in the full snapshots. For each snapshot, we selected galaxies with $\log (M_*/\mathrm{M_\odot}) > 11.5$ and $\log(sSFR) < -11$, following selection criteria consistent with those we adopted for the BOSS and DESI samples. Moreover, we derived light-weighted age, [Fe/H] and [$\alpha$/Fe] within one effective radius of each galaxy, to ensure consistency between simulated and observed values. The sMILES and TMJ models consider different elements in the definition of [$\alpha$/Fe] therefore we derived two estimates from the TNG element abundances. Indeed, the simulation provides individual abundances for H, He, C, N, O, Ne, Mg, Si, and Fe, but not for S, Ca, or Ti. However, the absence of these elements has a negligible impact on our analysis, as the dominant contribution to the total metal budget comes from O, C, N, and Mg \citep[see][for details]{labarbera2025}. For this reason, we computed [$\alpha$/Fe]$^\text{TNG}_\text{sMILES}$ using O, Ne, Mg, Si as $\alpha$-elements, while [$\alpha$/Fe]$^\text{TNG}_\text{TMJ}$ considering C, N, O, Ne, Mg and Si. 
We normalised the abundances using the solar reference value from \cite{asplund2009} ([Fe/H]$_\odot= -2.76$, [$\alpha$/Fe]$_{\odot,\text{sMILES}} = 0.81$, [$\alpha$/Fe]$_{\odot,\text{TMJ}} = 0.94$), since both TMJ and sMILES models express abundances in terms of logarithmic ratios relative to their adopted solar reference values.

\begin{figure}
   \centering
   \includegraphics[width=0.5\textwidth,height=0.8\textheight,keepaspectratio]{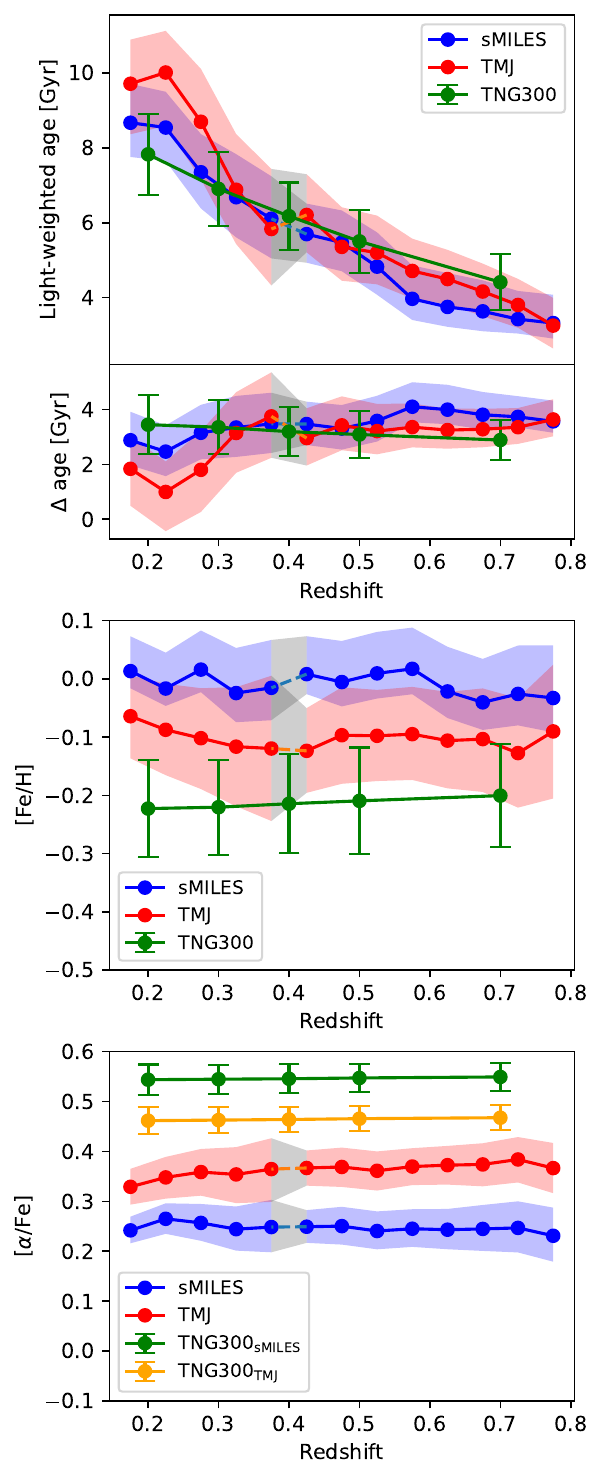}
      \caption{Evolution of the light-weighted age, [Fe/H] and [$\alpha$/Fe] in the redshift range $0.15<z<0.8$. The top panel includes a sub-panel showing the difference between the age of the Universe and the light-weighted age. The blue (red) data points represent the median value in each redshift bin using sMILES (TMJ) models, the shaded regions cover the typical uncertainties of the stacked spectra. The green data points and error bars are the median values and the dispersion of the stellar population parameters of the selected galaxies in five TNG$300$ snapshots. In the bottom panel the green (yellow) data points and error bars are the median values and the dispersion of the [$\alpha$/Fe] from TNG galaxies considering the normalisation given by the $\alpha$-abundances of sMILES (TMJ) models. The dashed black line in the top panel represents the age of the Universe at each redshift.
              }
         \label{fig:resultsbossdesi_tng}
         
   \end{figure}

Figure~\ref{fig:resultsbossdesi_tng} shows the the evolution of the light-weighted age, [Fe/H] and [$\alpha$/Fe] as a function of redshift for stacks of galaxies in the BOSS and DESI samples, alongside the parameters of the selected galaxies across the five TNG snapshots. The light-weighted age of TNG galaxies closely track the observed trends in the BOSS and DESI stacks, in both sMILES and TMJ analyses, with ages increasing from $4$ Gyr to $9$ Gyr toward lower redshifts, with an almost constant difference between the age of the Universe and the measured light-weighted age. This agreement lends support to the adopted selection of massive and quiescent galaxies in the TNG snapshots for the comparison.
Both the observed and simulated galaxies show nearly flat trends in [Fe/H] and [$\alpha$/Fe] toward lower redshift.  However, we obtained different absolute values for all the stellar population parameters. In particular, simulated galaxies show lower values of [Fe/H] (difference of $\sim 0.2$ dex with sMILES, $\sim 0.1$ with TMJ) and higher values of [$\alpha$/Fe] (difference of $\sim 0.3$ dex with sMILES, $\sim 0.1$ dex with TMJ). 

\section{Discussion}
The study of the age, [Fe/H] and the [$\alpha$/Fe] abundance ratio of the most massive quiescent galaxies across the cosmic time provides valuable insights into the framework of galaxy formation and evolution. Thanks to the extensive spectroscopic datasets from the BOSS and DESI surveys, we have been able to investigate how the most massive and quiescent galaxy population evolves in the redshift range $0.15<z<0.8$. Moreover, thanks to the larger mass dynamic range of the DESI sample, we studied these quantities in two mass bins. 

The DESI results reveal very consistent evolutionary trends for galaxies in the two mass bins, with the more massive systems showing slightly older ages, higher stellar metallicities, and enhanced [$\alpha$/Fe] ratios across the redshift range $0.4<z<0.8$. Although these systematic differences are within the $1\sigma$ uncertainties, they are suggestive of a mild downsizing effect, in which the more massive galaxies formed their stars earlier and over shorter timescales than their lower-mass counterparts \citep[e.g.][]{thomas2005epochs,gallazzi2005ages,fontanot2009}. The measured intrinsic scatter is smaller than, or at most comparable to the uncertainties across both mass and redshift bins for all parameters, suggesting a homogeneous stellar population within the DESI stacked spectra.

As shown in Figure~\ref{fig:resultsbossdesi}, the most massive quiescent galaxies exhibit a passive evolution of the light-weighted age, with an almost constant offset from the Universe age. 
Our measurements also reveal no significant evolution in the global stellar metallicity of these most massive quiescent galaxies from $z \sim 0.8$ to $z \sim 0.15$, supporting the picture of their passive evolution.

A comparison between the sMILES and TMJ models shows that the derived stellar population parameters follow the same evolution trends in both cases. In particular, the [Fe/H] measurements using TMJ are systematically higher of around $0.1$ dex with respect to sMILES results. 
For [$\alpha$/Fe], our estimates effectively trace [Mg/Fe], as Mg is the dominant 
$\alpha$-element feature driving the fit. Nonetheless, the absolute 
[$\alpha$/Fe] values differ between sMILES and TMJ of around $0.1$ dex. These differences are expected, since the two models rely on different chemical prescriptions and SSP construction. In particular, their distinct definitions of the $\alpha$-element mixture naturally produce slightly different [Fe/H] and [$\alpha$/Fe] values. The light-weighted ages derived with the two models are consistent with each other, despite the different stellar evolutionary prescription adopted. Indeed sMILES rely on BaSTI isochrones, while TMJ uses \cite{cassisi1997}, as well as additional model-specific recipes for the treatment of horizontal-branch and AGB phases.
Despite these offsets in the absolute values of [Fe/H] and [$\alpha$/Fe], the consistency of the evolutionary trends across the two model sets indicates that our main conclusions are robust to the choice of stellar population model. We can then proceed to compare these results with the predictions from cosmological simulations to assess whether current galaxy formation models reproduce the observed trends. Regardless of the absolute offset, the redshift evolution of [$\alpha$/Fe] in TNG galaxies closely follows the observed trend obtained from the observations, both showing a constant trend across the entire redshift range. However, the [$\alpha$/Fe] values predicted by TNG simulations for massive quiescent galaxies are systematically higher by $\sim 0.2$ dex than those derived from observations \citep{naiman2018,kim2023}. This result is confirmed by our results, for which we have a difference of $\sim 0.3$ dex from sMILES results, and $\sim 0.1$ dex from the TMJ ones. This discrepancy likely originates from differences in nucleosynthetic yields and stellar population modelling. Indeed, according to \cite{naiman2018}, the offset may point to an overproduction of $\alpha$-elements relative to Fe from core-collapse supernovae or a too low rate of Type Ia supernovae in the simulation physics. Regarding the [Fe/H] abundance patterns, previous works report broadly consistent or slightly higher values in the highest-mass bin in TNG than in their observational sample \citep{naiman2018,kim2023}. However, these studies rely on small observed datasets, especially at high mass, making the comparison sensitive to the sample selection and aperture effects. With our larger and more homogeneous sample, instead we find that the observed [Fe/H] is higher (by $\sim0.2$ dex with sMILES, $\sim 0.1$ dex with TMJ) than the one of the simulated galaxies. Given the rapid quenching of the TNG galaxies, most stars form before the bulk of Type Ia supernovae explode, therefore much of the late-time iron enrichment does not get incorporated into new stars. The simulations therefore naturally predict slightly lower stellar [Fe/H] than observed.

Given the wide redshift range covered by our analysis, we can compare our results with previous studies of massive quiescent galaxies at different epochs.
Numerous studies, based on different data and methodologies, consistently support a picture in which the highest mass quiescent galaxies evolve passively over the last $8$ Gyr. Indeed, \cite{gallazzi2014} found a strong similarity in the properties of quiescent galaxies at $z \sim 0.7$ and local quiescent systems for $\log (M_*/\mathrm{M_\odot}) > 11.5$, once accounting for passive evolution. Other works based on LEGA-C data \citep{beverage2023,bevacqua2024}, although limited to smaller samples, found that the formation epochs of the most massive galaxies at $0.6 < z < 0.75$ are consistent with those inferred for local galaxies. Stacking analyses at different redshifts further confirm the passive evolution of high–mass quiescent galaxies \citep{choi2014assembly}. 

While the total stellar metallicity traces the overall metal content accumulated during star formation, the [$\alpha$/Fe] abundance ratio provides the complementary information on the timescales of that enrichment \citep[][]{thomas1999}. A key emerging result from both our results and simulations is that [$\alpha$/Fe] shows no evolution across cosmic time for massive quiescent galaxies. \cite{choi2014assembly} and \cite{Leethochawalit2019}, analysing composite spectra of massive quiescent galaxies up to $z \sim 0.5$, find very weak or no evolution in [Mg/Fe] compared to the local Universe. Similarly, LEGA-C massive quiescent galaxies at $z \sim 0.7$ show [Mg/Fe] ratios comparable to those of local massive early-type systems \citep{beverage2021,bevacqua2023}. These results indicate that the most massive quiescent galaxies observed at $0.15<z<0.8$ had already completed their star formation by $z\gtrsim1.5$–$2$. Their uniformly high [Mg/Fe] ratios across this redshift range suggest that both the galaxies observed at $z\sim0.8$ and those at $z\sim0.15$ formed and quenched their stars at similar early epochs, implying rapid formation timescales followed by passive evolution.

At higher redshift ($z \sim1 - 1.3$), \cite{carnall2019,carnall2022} reported slightly lower metallicities for galaxies with $\log (M_*/\mathrm{M_\odot}) > 11$, with an offset of $ \sim 0.2-0.3$ dex below the local values. Their studies also found [Mg/Fe] ratios in massive quiescent galaxies comparable to those of local early-type systems, suggesting that the $\alpha$-enhancement, and thus the short formation timescales, were already established by $z \sim 1$. However, their stellar metallicities results are consistent with an earlier evolutionary stage of the quiescent population. Between $z\sim 1.3 $ and $z\sim 0.8$, that correspond to around $2$ Gyr of cosmic time, the number density of quiescent galaxies with $\log (M_*/\mathrm{M_\odot}) > 11.5$ increases by a factor of $\sim 4-5$ \citep{muzzin2013}. This implies that a large number of quenched systems joined the massive quiescent population during this period. These additions have been probably slightly more metal-rich but shared similarly high [Mg/Fe] ratios, producing a mild apparent evolution in metallicity while maintaining a roughly constant $\alpha$-enhancement. This behaviour is indicative of a classical progenitor bias, where the growth of the quiescent population at lower redshift reflects the addition of recently quenched galaxies rather than the intrinsic evolution of the original quiescent systems. 
Further support for this interpretation comes from \cite{mendel2015}, who measured stellar population ages of massive quiescent galaxies at $1.5 < z < 2$. Their results, when compared with other studies at lower redshift, indicate that these high-redshift quiescent galaxies do not represent the bulk of the quiescent population observed at $z < 1$, but rather an earlier generation of massive systems that quenched rapidly and evolved passively thereafter. At $z < 1$, the stellar ages of massive quiescent galaxies are consistent with passive evolution, whereas at $1 < z < 2$ they appear to saturate to a young ($\sim 1$ Gyr) age, reflecting the changing demographics of the average progenitor population.
In contrast, from $z \sim 0.8$ to $z \sim 0.15$ our results show that the stellar population properties of the most massive quiescent galaxies remain remarkably constant, despite the nearly $5$ Gyr of cosmic time and an additional increase in number density by a factor of $3-4$ \citep[e.g.,][]{muzzin2013, donnari2021}. Our observational results then demonstrate that below $z \sim 0.8$ the progenitor bias is no longer significant, and the evolution of the massive quiescent population is largely explained by a simple, passively evolving, scenario. The galaxies added at later times were already quenched, having formed most of their stars at earlier epochs. Therefore, at $z < 0.8$, their evolution likely involved mostly minor dry mergers, which increased their stellar mass and size without significantly affecting the light-weighted age, [M/H] or [$\alpha$/Fe] distribution. Consequently, while the build-up of the quiescent population between $z\sim2$ and $z\sim0.8$ is shaped by progenitor bias, at later times the most massive quiescent galaxies evolve passively, with structural growth dominating over further chemical or star-formation evolution. These findings naturally connect with studies of the local early-type galaxy population. In particular, \cite{thomas2005epochs} inferred that the formation of the total stellar mass in massive early-type was already nearly complete by $z \sim 1$, with subsequent stellar mass growth limited to low-mass systems. Our results directly confirm this prediction, painint a coherent picture in which the most massive quiescent galaxies evolve predominantly through passive aging and minor structural transformations.

\section{Summary and conclusions}
\label{sec:conclusion}
The analysis presented here provides the first homogeneous and continuous mapping of the stellar population properties of the most massive quiescent galaxies ($\log (M_*/\mathrm{M_\odot})> 11.5$) across the last $\sim7$ Gyr of cosmic time ($0.15 < z < 0.8$). This work has been made possible by the combination of the BOSS and DESI surveys, whose unprecedented spectral coverage and large sample size provide the foundation for our analysis. By applying a homogeneous selection across the two datasets we were able to trace the evolution of the most massive quiescent systems across a wide redshift range. We measured trends of [Fe/H], [$\alpha$/Fe], and light–weighted age of the most massive quiescent galaxies as continuous functions of redshift, and consistently fitting the same set of spectral indices at all redshifts. This methodological homogeneity removes inter-survey calibration biases that have so far limited direct evolutionary comparisons.

The resulting trends show no significant evolution in [$\alpha$/Fe] and [Fe/H], and an age increase fully consistent with passive stellar evolution. These results are robust against the choice of stellar population models and analysis assumptions, and they are in excellent agreement with the predictions from IllustrisTNG, which similarly predict negligible evolution in the chemical properties of the most massive quiescent galaxies at $z \le 0.8$.
This remarkable stability indicates that, below $z\sim0.8$, the evolution of the most massive quiescent population is genuinely passive. Galaxies that join the population at later times already exhibit stellar properties similar to those of systems in place at $z\sim0.8$, suggesting that they formed the bulk of their stars at similarly early epochs. This implies that, below $z \sim 0.8$, the growth of the most massive quiescent galaxies is dominated by passive mass assembly through minor dry mergers rather than by the addition of freshly quenched systems.

Beyond confirming the qualitative expectations from previous snapshots and cosmological simulations, this work provides a quantitative characterisation of the passive–evolution regime in a way that was previously unattainable. Indeed, it provides a quantitative benchmark for the passive evolution of the most massive galaxies, offering a solid empirical basis for future comparisons with hydrodynamical simulations and chemical–enrichment models.
Forthcoming spectroscopic surveys will enable a further step forward by extending this analysis to individual galaxies and exploring the dispersion of stellar population properties as a function of stellar mass, environment, and cosmic time.
Upcoming high–multiplex, wide–field instruments such as WEAVE and 4MOST facilities \citep[][]{dejong2019,jin2024}   will provide spectra for thousands of galaxies over wide spectral ranges and redshift intervals comparable to those of BOSS and DESI, but with signal–to–noise ratios high enough to derive detailed stellar population parameters on a galaxy–by–galaxy basis \citep[e.g. ,4MOST-StePS and WEAVE-StePS][]{iovino2023,iovino2023Msn}.
Moreover, the survey MOONRISE \citep{maiolino2020Msn} will exploit the unique observing capabilities of MOONS at VLT to provide the data to extend this approach to $z \ge 1$, providing a direct view of the earliest phases of quenching and completing the empirical picture of massive galaxy evolution across cosmic time.

\section*{Acknowledgements}
FLB acknowledges support from INAF 2023 RSN1 Minigrant 1.05.23.04.01.

\section*{Data Availability}
The observational data used in this paper are publicly accessible. BOSS data can be obtained from the SDSS Data Release Server (\url{https://data.sdss.org/sas/}), DESI data are available through the DESI Public Data Release portal(\url{https://data.desi.lbl.gov/doc/}), and IllustrisTNG data can be found at \url{https://www.tng-project.org/data/downloads/TNG300-1/}. The models and code used in this work rely on publicly available resources cited in the manuscript. No proprietary data were used.
 



\bibliographystyle{mnras}
\bibliography{example} 

@ARTICLE{nelson2019,
       author = {{Nelson}, Dylan and {Springel}, Volker and {Pillepich}, Annalisa and {Rodriguez-Gomez}, Vicente and {Torrey}, Paul and {Genel}, Shy and {Vogelsberger}, Mark and {Pakmor}, Ruediger and {Marinacci}, Federico and {Weinberger}, Rainer and {Kelley}, Luke and {Lovell}, Mark and {Diemer}, Benedikt and {Hernquist}, Lars},
        title = "{The IllustrisTNG simulations: public data release}",
      journal = {Computational Astrophysics and Cosmology},
         year = 2019,
        month = may,
       volume = {6},
       number = {1},
          eid = {2},
        pages = {2},
          doi = {10.1186/s40668-019-0028-x},
archivePrefix = {arXiv},
       eprint = {1812.05609},
 primaryClass = {astro-ph.GA},
       adsurl = {https://ui.adsabs.harvard.edu/abs/2019ComAC...6....2N}
}

@ARTICLE{beverage2021,
       author = {{Beverage}, Aliza G. and {Kriek}, Mariska and {Conroy}, Charlie and {Bezanson}, Rachel and {Franx}, Marijn and {van der Wel}, Arjen},
        title = "{Elemental Abundances and Ages of z   0.7 Quiescent Galaxies on the Mass-Size Plane: Implication for Chemical Enrichment and Star Formation Quenching}",
      journal = {\apjl},
         year = 2021,
        month = aug,
       volume = {917},
       number = {1},
          eid = {L1},
        pages = {L1},
          doi = {10.3847/2041-8213/ac12cd},
archivePrefix = {arXiv},
       eprint = {2105.12750},
 primaryClass = {astro-ph.GA},
       adsurl = {https://ui.adsabs.harvard.edu/abs/2021ApJ...917L...1B}
}

@ARTICLE{Leethochawalit2019,
       author = {{Leethochawalit}, Nicha and {Kirby}, Evan N. and {Ellis}, Richard S. and {Moran}, Sean M. and {Treu}, Tommaso},
        title = "{Evolution of the Stellar Mass-Metallicity Relation. II. Constraints on Galactic Outflows from the Mg Abundances of Quiescent Galaxies}",
      journal = {\apj},
         year = 2019,
        month = nov,
       volume = {885},
       number = {2},
          eid = {100},
        pages = {100},
          doi = {10.3847/1538-4357/ab4809},
archivePrefix = {arXiv},
       eprint = {1909.11680},
 primaryClass = {astro-ph.GA},
       adsurl = {https://ui.adsabs.harvard.edu/abs/2019ApJ...885..100L}
}

@ARTICLE{greggio1983,
       author = {{Greggio}, L. and {Renzini}, A.},
        title = "{The binary model for type I supernovae - Theoretical rates}",
      journal = {\aap},
         year = 1983,
        month = feb,
       volume = {118},
       number = {2},
        pages = {217-222},
       adsurl = {https://ui.adsabs.harvard.edu/abs/1983A&A...118..217G}
}

@ARTICLE{carnall2022,
       author = {{Carnall}, Adam C. and {McLure}, Ross J. and {Dunlop}, James S. and {Hamadouche}, Massissilia and {Cullen}, Fergus and {McLeod}, Derek J. and {Begley}, Ryan and {Amorin}, Ricardo and {Bolzonella}, Micol and {Castellano}, Marco and {Cimatti}, Andrea and {Fontanot}, Fabio and {Gargiulo}, Adriana and {Garilli}, Bianca and {Mannucci}, Filippo and {Pentericci}, Laura and {Talia}, Margherita and {Zamorani}, Giovani and {Calabro}, Antonello and {Cresci}, Giovanni and {Hathi}, Nimish P.},
        title = "{The Stellar Metallicities of Massive Quiescent Galaxies at 1.0 < z < 1.3 from KMOS + VANDELS}",
      journal = {\apj},
         year = 2022,
        month = apr,
       volume = {929},
       number = {2},
          eid = {131},
        pages = {131},
          doi = {10.3847/1538-4357/ac5b62},
archivePrefix = {arXiv},
       eprint = {2108.13430},
 primaryClass = {astro-ph.GA},
       adsurl = {https://ui.adsabs.harvard.edu/abs/2022ApJ...929..131C}
}

@ARTICLE{carnall2019,
       author = {{Carnall}, A.~C. and {McLure}, R.~J. and {Dunlop}, J.~S. and {Cullen}, F. and {McLeod}, D.~J. and {Wild}, V. and {Johnson}, B.~D. and {Appleby}, S. and {Dav{\'e}}, R. and {Amorin}, R. and {Bolzonella}, M. and {Castellano}, M. and {Cimatti}, A. and {Cucciati}, O. and {Gargiulo}, A. and {Garilli}, B. and {Marchi}, F. and {Pentericci}, L. and {Pozzetti}, L. and {Schreiber}, C. and {Talia}, M. and {Zamorani}, G.},
        title = "{The VANDELS survey: the star-formation histories of massive quiescent galaxies at 1.0 < z < 1.3}",
      journal = {\mnras},
         year = 2019,
        month = nov,
       volume = {490},
       number = {1},
        pages = {417-439},
          doi = {10.1093/mnras/stz2544},
archivePrefix = {arXiv},
       eprint = {1903.11082},
 primaryClass = {astro-ph.GA},
       adsurl = {https://ui.adsabs.harvard.edu/abs/2019MNRAS.490..417C}
}

@ARTICLE{zibetti2020,
       author = {{Zibetti}, Stefano and {Gallazzi}, Anna R. and {Hirschmann}, Michaela and {Consolandi}, Guido and {Falc{\'o}n-Barroso}, Jes{\'u}s and {van de Ven}, Glenn and {Lyubenova}, Mariya},
        title = "{Insights into formation scenarios of massive early-type galaxies from spatially resolved stellar population analysis in CALIFA}",
      journal = {\mnras},
         year = 2020,
        month = jan,
       volume = {491},
       number = {3},
        pages = {3562-3585},
          doi = {10.1093/mnras/stz3205},
archivePrefix = {arXiv},
       eprint = {1906.02209},
 primaryClass = {astro-ph.GA},
       adsurl = {https://ui.adsabs.harvard.edu/abs/2020MNRAS.491.3562Z}
}

@ARTICLE{santucci2020,
       author = {{Santucci}, Giulia and {Brough}, Sarah and {Scott}, Nicholas and {Montes}, Mireia and {Owers}, Matt S. and {van Sande}, Jesse and {Bland-Hawthorn}, Joss and {Bryant}, Julia J. and {Croom}, Scott M. and {Ferreras}, Ignacio and {Lawrence}, Jon S. and {L{\'o}pez-S{\'a}nchez}, {\'A}ngel R. and {Richards}, Samuel N.},
        title = "{The SAMI Galaxy Survey: Stellar Population Gradients of Central Galaxies}",
      journal = {\apj},
         year = 2020,
        month = jun,
       volume = {896},
       number = {1},
          eid = {75},
        pages = {75},
          doi = {10.3847/1538-4357/ab92a9},
archivePrefix = {arXiv},
       eprint = {2005.00541},
 primaryClass = {astro-ph.GA},
       adsurl = {https://ui.adsabs.harvard.edu/abs/2020ApJ...896...75S}
}

@ARTICLE{parikh2021,
       author = {{Parikh}, Taniya and {Thomas}, Daniel and {Maraston}, Claudia and {Westfall}, Kyle B. and {Andrews}, Brett H. and {Boardman}, Nicholas Fraser and {Drory}, Niv and {Oyarzun}, Grecco},
        title = "{SDSS-IV MaNGA: radial gradients in stellar population properties of early-type and late-type galaxies}",
      journal = {\mnras},
         year = 2021,
        month = apr,
       volume = {502},
       number = {4},
        pages = {5508-5527},
          doi = {10.1093/mnras/stab449},
archivePrefix = {arXiv},
       eprint = {2102.06703},
 primaryClass = {astro-ph.GA},
       adsurl = {https://ui.adsabs.harvard.edu/abs/2021MNRAS.502.5508P}
}

@ARTICLE{maiolino2020Msn,
       author = {{Maiolino}, R. and {Cirasuolo}, M. and {Afonso}, J. and {Bauer}, F.~E. and {Bowler}, R. and {Cucciati}, O. and {Daddi}, E. and {De Lucia}, G. and {Evans}, C. and {Flores}, H. and {Gargiulo}, A. and {Garilli}, B. and {Jablonka}, P. and {Jarvis}, M. and {Kneib}, J.-P. and {Lilly}, S. and {Looser}, T. and {Magliocchetti}, M. and {Man}, Z. and {Mannucci}, F. and {Maurogordato}, S. and {McLure}, R.~J. and {Norberg}, P. and {Oesch}, P. and {Oliva}, E. and {Paltani}, S. and {Pappalardo}, C. and {Peng}, Y. and {Pentericci}, L. and {Pozzetti}, L. and {Renzini}, A. and {Rodrigues}, M. and {Royer}, F. and {Serjeant}, S. and {Vanzi}, L. and {Wild}, V. and {Zamorani}, G.},
        title = "{MOONRISE: The Main MOONS GTO Extragalactic Survey}",
      journal = {The Messenger},
         year = 2020,
        month = jun,
       volume = {180},
        pages = {24-29},
          doi = {10.18727/0722-6691/5197},
archivePrefix = {arXiv},
       eprint = {2009.00644},
 primaryClass = {astro-ph.GA},
       adsurl = {https://ui.adsabs.harvard.edu/abs/2020Msngr.180...24M}
}

@ARTICLE{mendel2015,
       author = {{Mendel}, J. Trevor and {Saglia}, Roberto P. and {Bender}, Ralf and {Beifiori}, Alessandra and {Chan}, Jeffrey and {Fossati}, Matteo and {Wilman}, David J. and {Bandara}, Kaushala and {Brammer}, Gabriel B. and {F{\"o}rster Schreiber}, Natascha M. and {Galametz}, Audrey and {Kulkarni}, Sandesh and {Momcheva}, Ivelina G. and {Nelson}, Erica J. and {van Dokkum}, Pieter G. and {Whitaker}, Katherine E. and {Wuyts}, Stijn},
        title = "{First Results from the VIRIAL Survey: The Stellar Content of UVJ-selected Quiescent Galaxies at 1.5 < z < 2 from KMOS}",
      journal = {\apjl},
         year = 2015,
        month = may,
       volume = {804},
       number = {1},
          eid = {L4},
        pages = {L4},
          doi = {10.1088/2041-8205/804/1/L4},
archivePrefix = {arXiv},
       eprint = {1503.08831},
 primaryClass = {astro-ph.GA},
       adsurl = {https://ui.adsabs.harvard.edu/abs/2015ApJ...804L...4M}
}

@ARTICLE{jafariyazani2025,
       author = {{Jafariyazani}, Marziye and {Newman}, Andrew B. and {Mobasher}, Bahram and {Belli}, Sirio and {Ellis}, Richard S. and {Faisst}, Andreas L.},
        title = "{Chemical Abundances of Early Quiescent Galaxies: New Observations and Modeling Impacts}",
      journal = {\apj},
         year = 2025,
        month = jun,
       volume = {986},
       number = {2},
          eid = {148},
        pages = {148},
          doi = {10.3847/1538-4357/addbdc},
archivePrefix = {arXiv},
       eprint = {2406.03549},
 primaryClass = {astro-ph.GA},
       adsurl = {https://ui.adsabs.harvard.edu/abs/2025ApJ...986..148J}
}

@ARTICLE{bevacqua2024,
       author = {{Bevacqua}, Davide and {Saracco}, Paolo and {Boecker}, Alina and {D'Ago}, Giuseppe and {De Lucia}, Gabriella and {De Propris}, Roberto and {La Barbera}, Francesco and {Pasquali}, Anna and {Spiniello}, Chiara and {Tortora}, Crescenzo},
        title = "{A new perspective on the stellar mass-metallicity relation of quiescent galaxies from the LEGA-C survey}",
      journal = {\aap},
         year = 2024,
        month = oct,
       volume = {690},
          eid = {A150},
        pages = {A150},
          doi = {10.1051/0004-6361/202348831},
archivePrefix = {arXiv},
       eprint = {2407.12704},
 primaryClass = {astro-ph.GA},
       adsurl = {https://ui.adsabs.harvard.edu/abs/2024A&A...690A.150B}
}

@ARTICLE{cassisi1997,
       author = {{Cassisi}, S. and {Castellani}, M. and {Castellani}, V.},
        title = "{Intermediate-age metal deficient stellar populations: the case of metallicity Z=0.00001.}",
      journal = {\aap},
         year = 1997,
        month = jan,
       volume = {317},
        pages = {108-113},
          doi = {10.48550/arXiv.astro-ph/9603023},
archivePrefix = {arXiv},
       eprint = {astro-ph/9603023},
 primaryClass = {astro-ph},
       adsurl = {https://ui.adsabs.harvard.edu/abs/1997A&A...317..108C}
}

@ARTICLE{thomas2011,
       author = {{Thomas}, Daniel and {Maraston}, Claudia and {Johansson}, Jonas},
        title = "{Flux-calibrated stellar population models of Lick absorption-line indices with variable element abundance ratios}",
      journal = {\mnras},
         year = 2011,
        month = apr,
       volume = {412},
       number = {4},
        pages = {2183-2198},
          doi = {10.1111/j.1365-2966.2010.18049.x},
archivePrefix = {arXiv},
       eprint = {1010.4569},
 primaryClass = {astro-ph.CO},
       adsurl = {https://ui.adsabs.harvard.edu/abs/2011MNRAS.412.2183T}
}

@ARTICLE{matteucci1987,
       author = {{Matteucci}, F. and {Tornambe}, A.},
        title = "{Chemical evolution of elliptical galaxies}",
      journal = {\aap},
         year = 1987,
        month = oct,
       volume = {185},
       number = {1-2},
        pages = {51-60},
       adsurl = {https://ui.adsabs.harvard.edu/abs/1987A&A...185...51M}
}

@ARTICLE{thomas1999,
       author = {{Thomas}, D. and {Greggio}, L. and {Bender}, R.},
        title = "{Constraints on galaxy formation from alpha-enhancement in luminous elliptical galaxies}",
      journal = {\mnras},
         year = 1999,
        month = jan,
       volume = {302},
       number = {3},
        pages = {537-548},
          doi = {10.1046/j.1365-8711.1999.02138.x},
archivePrefix = {arXiv},
       eprint = {astro-ph/9809261},
 primaryClass = {astro-ph},
       adsurl = {https://ui.adsabs.harvard.edu/abs/1999MNRAS.302..537T}
}

@ARTICLE{tinsley1979,
       author = {{Tinsley}, B.~M.},
        title = "{Stellar lifetimes and abundance ratios in chemical evolution.}",
      journal = {\apj},
         year = 1979,
        month = may,
       volume = {229},
        pages = {1046-1056},
          doi = {10.1086/157039},
       adsurl = {https://ui.adsabs.harvard.edu/abs/1979ApJ...229.1046T}
}

@ARTICLE{renzini2006,
       author = {{Renzini}, Alvio},
        title = "{Stellar Population Diagnostics of Elliptical Galaxy Formation}",
      journal = {\araa},
         year = 2006,
        month = sep,
       volume = {44},
       number = {1},
        pages = {141-192},
          doi = {10.1146/annurev.astro.44.051905.092450},
archivePrefix = {arXiv},
       eprint = {astro-ph/0603479},
 primaryClass = {astro-ph},
       adsurl = {https://ui.adsabs.harvard.edu/abs/2006ARA&A..44..141R}
}

@ARTICLE{favole2018,
       author = {{Favole}, Ginevra and {Montero-Dorta}, Antonio D. and {Prada}, Francisco and {Rodr{\'\i}guez-Torres}, Sergio A. and {Schlegel}, David J.},
        title = "{The mass-size relation of luminous red galaxies from BOSS and DECaLS}",
      journal = {\mnras},
         year = 2018,
        month = oct,
       volume = {480},
       number = {1},
        pages = {1415-1425},
          doi = {10.1093/mnras/sty1947},
archivePrefix = {arXiv},
       eprint = {1802.01596},
 primaryClass = {astro-ph.GA},
       adsurl = {https://ui.adsabs.harvard.edu/abs/2018MNRAS.480.1415F}
}

@ARTICLE{naiman2018,
       author = {{Naiman}, Jill P. and {Pillepich}, Annalisa and {Springel}, Volker and {Ramirez-Ruiz}, Enrico and {Torrey}, Paul and {Vogelsberger}, Mark and {Pakmor}, R{\"u}diger and {Nelson}, Dylan and {Marinacci}, Federico and {Hernquist}, Lars and {Weinberger}, Rainer and {Genel}, Shy},
        title = "{First results from the IllustrisTNG simulations: a tale of two elements - chemical evolution of magnesium and europium}",
      journal = {\mnras},
         year = 2018,
        month = jun,
       volume = {477},
       number = {1},
        pages = {1206-1224},
          doi = {10.1093/mnras/sty618},
archivePrefix = {arXiv},
       eprint = {1707.03401},
 primaryClass = {astro-ph.GA},
       adsurl = {https://ui.adsabs.harvard.edu/abs/2018MNRAS.477.1206N}
}

@ARTICLE{desi2016a,
       author = {{DESI Collaboration} and {Aghamousa}, Amir and {Aguilar}, Jessica and {Ahlen}, Steve and {Alam}, Shadab and {Allen}, Lori E. and {Allende Prieto}, Carlos and {Annis}, James and {Bailey}, Stephen and {Balland}, Christophe and {Ballester}, Otger and {Baltay}, Charles and {Beaufore}, Lucas and {Bebek}, Chris and {Beers}, Timothy C. and {Bell}, Eric F. and {Bernal}, Jos{\'e} Luis and {Besuner}, Robert and {Beutler}, Florian and {Blake}, Chris and {Bleuler}, Hannes and {Blomqvist}, Michael and {Blum}, Robert and {Bolton}, Adam S. and {Briceno}, Cesar and {Brooks}, David and {Brownstein}, Joel R. and {Buckley-Geer}, Elizabeth and {Burden}, Angela and {Burtin}, Etienne and {Busca}, Nicolas G. and {Cahn}, Robert N. and {Cai}, Yan-Chuan and {Cardiel-Sas}, Laia and {Carlberg}, Raymond G. and {Carton}, Pierre-Henri and {Casas}, Ricard and {Castander}, Francisco J. and {Cervantes-Cota}, Jorge L. and {Claybaugh}, Todd M. and {Close}, Madeline and {Coker}, Carl T. and {Cole}, Shaun and {Comparat}, Johan and {Cooper}, Andrew P. and {Cousinou}, M. -C. and {Crocce}, Martin and {Cuby}, Jean-Gabriel and {Cunningham}, Daniel P. and {Davis}, Tamara M. and {Dawson}, Kyle S. and {de la Macorra}, Axel and {De Vicente}, Juan and {Delubac}, Timoth{\'e}e and {Derwent}, Mark and {Dey}, Arjun and {Dhungana}, Govinda and {Ding}, Zhejie and {Doel}, Peter and {Duan}, Yutong T. and {Ealet}, Anne and {Edelstein}, Jerry and {Eftekharzadeh}, Sarah and {Eisenstein}, Daniel J. and {Elliott}, Ann and {Escoffier}, St{\'e}phanie and {Evatt}, Matthew and {Fagrelius}, Parker and {Fan}, Xiaohui and {Fanning}, Kevin and {Farahi}, Arya and {Farihi}, Jay and {Favole}, Ginevra and {Feng}, Yu and {Fernandez}, Enrique and {Findlay}, Joseph R. and {Finkbeiner}, Douglas P. and {Fitzpatrick}, Michael J. and {Flaugher}, Brenna and {Flender}, Samuel and {Font-Ribera}, Andreu and {Forero-Romero}, Jaime E. and {Fosalba}, Pablo and {Frenk}, Carlos S. and {Fumagalli}, Michele and {Gaensicke}, Boris T. and {Gallo}, Giuseppe and {Garcia-Bellido}, Juan and {Gaztanaga}, Enrique and {Pietro Gentile Fusillo}, Nicola and {Gerard}, Terry and {Gershkovich}, Irena and {Giannantonio}, Tommaso and {Gillet}, Denis and {Gonzalez-de-Rivera}, Guillermo and {Gonzalez-Perez}, Violeta and {Gott}, Shelby and {Graur}, Or and {Gutierrez}, Gaston and {Guy}, Julien and {Habib}, Salman and {Heetderks}, Henry and {Heetderks}, Ian and {Heitmann}, Katrin and {Hellwing}, Wojciech A. and {Herrera}, David A. and {Ho}, Shirley and {Holland}, Stephen and {Honscheid}, Klaus and {Huff}, Eric and {Hutchinson}, Timothy A. and {Huterer}, Dragan and {Hwang}, Ho Seong and {Illa Laguna}, Joseph Maria and {Ishikawa}, Yuzo and {Jacobs}, Dianna and {Jeffrey}, Niall and {Jelinsky}, Patrick and {Jennings}, Elise and {Jiang}, Linhua and {Jimenez}, Jorge and {Johnson}, Jennifer and {Joyce}, Richard and {Jullo}, Eric and {Juneau}, St{\'e}phanie and {Kama}, Sami and {Karcher}, Armin and {Karkar}, Sonia and {Kehoe}, Robert and {Kennamer}, Noble and {Kent}, Stephen and {Kilbinger}, Martin and {Kim}, Alex G. and {Kirkby}, David and {Kisner}, Theodore and {Kitanidis}, Ellie and {Kneib}, Jean-Paul and {Koposov}, Sergey and {Kovacs}, Eve and {Koyama}, Kazuya and {Kremin}, Anthony and {Kron}, Richard and {Kronig}, Luzius and {Kueter-Young}, Andrea and {Lacey}, Cedric G. and {Lafever}, Robin and {Lahav}, Ofer and {Lambert}, Andrew and {Lampton}, Michael and {Landriau}, Martin and {Lang}, Dustin and {Lauer}, Tod R. and {Le Goff}, Jean-Marc and {Le Guillou}, Laurent and {Le Van Suu}, Auguste and {Lee}, Jae Hyeon and {Lee}, Su-Jeong and {Leitner}, Daniela and {Lesser}, Michael and {Levi}, Michael E. and {L'Huillier}, Benjamin and {Li}, Baojiu and {Liang}, Ming and {Lin}, Huan and {Linder}, Eric and {Loebman}, Sarah R. and {Luki{\'c}}, Zarija and {Ma}, Jun and {MacCrann}, Niall and {Magneville}, Christophe and {Makarem}, Laleh and {Manera}, Marc and {Manser}, Christopher J. and {Marshall}, Robert and {Martini}, Paul and {Massey}, Richard and {Matheson}, Thomas and {McCauley}, Jeremy and {McDonald}, Patrick and {McGreer}, Ian D. and {Meisner}, Aaron and {Metcalfe}, Nigel and {Miller}, Timothy N. and {Miquel}, Ramon and {Moustakas}, John and {Myers}, Adam and {Naik}, Milind and {Newman}, Jeffrey A. and {Nichol}, Robert C. and {Nicola}, Andrina and {Nicolati da Costa}, Luiz and {Nie}, Jundan and {Niz}, Gustavo and {Norberg}, Peder and {Nord}, Brian and {Norman}, Dara and {Nugent}, Peter and {O'Brien}, Thomas and {Oh}, Minji and {Olsen}, Knut A.~G.},
        title = "{The DESI Experiment Part I: Science,Targeting, and Survey Design}",
      journal = {arXiv e-prints},
         year = 2016,
        month = oct,
          eid = {arXiv:1611.00036},
        pages = {arXiv:1611.00036},
          doi = {10.48550/arXiv.1611.00036},
archivePrefix = {arXiv},
       eprint = {1611.00036},
 primaryClass = {astro-ph.IM},
       adsurl = {https://ui.adsabs.harvard.edu/abs/2016arXiv161100036D}
}

@ARTICLE{eisenstein2005,
       author = {{Eisenstein}, Daniel J. and {Zehavi}, Idit and {Hogg}, David W. and {Scoccimarro}, Roman and {Blanton}, Michael R. and {Nichol}, Robert C. and {Scranton}, Ryan and {Seo}, Hee-Jong and {Tegmark}, Max and {Zheng}, Zheng and {Anderson}, Scott F. and {Annis}, Jim and {Bahcall}, Neta and {Brinkmann}, Jon and {Burles}, Scott and {Castander}, Francisco J. and {Connolly}, Andrew and {Csabai}, Istvan and {Doi}, Mamoru and {Fukugita}, Masataka and {Frieman}, Joshua A. and {Glazebrook}, Karl and {Gunn}, James E. and {Hendry}, John S. and {Hennessy}, Gregory and {Ivezi{\'c}}, Zeljko and {Kent}, Stephen and {Knapp}, Gillian R. and {Lin}, Huan and {Loh}, Yeong-Shang and {Lupton}, Robert H. and {Margon}, Bruce and {McKay}, Timothy A. and {Meiksin}, Avery and {Munn}, Jeffery A. and {Pope}, Adrian and {Richmond}, Michael W. and {Schlegel}, David and {Schneider}, Donald P. and {Shimasaku}, Kazuhiro and {Stoughton}, Christopher and {Strauss}, Michael A. and {SubbaRao}, Mark and {Szalay}, Alexander S. and {Szapudi}, Istv{\'a}n and {Tucker}, Douglas L. and {Yanny}, Brian and {York}, Donald G.},
        title = "{Detection of the Baryon Acoustic Peak in the Large-Scale Correlation Function of SDSS Luminous Red Galaxies}",
      journal = {\apj},
         year = 2005,
        month = nov,
       volume = {633},
       number = {2},
        pages = {560-574},
          doi = {10.1086/466512},
archivePrefix = {arXiv},
       eprint = {astro-ph/0501171},
 primaryClass = {astro-ph},
       adsurl = {https://ui.adsabs.harvard.edu/abs/2005ApJ...633..560E}
}

@ARTICLE{eisenstein2001,
       author = {{Eisenstein}, Daniel J. and {Annis}, James and {Gunn}, James E. and {Szalay}, Alexander S. and {Connolly}, Andrew J. and {Nichol}, R.~C. and {Bahcall}, Neta A. and {Bernardi}, Mariangela and {Burles}, Scott and {Castander}, Francisco J. and {Fukugita}, Masataka and {Hogg}, David W. and {Ivezi{\'c}}, {\v{Z}}eljko and {Knapp}, G.~R. and {Lupton}, Robert H. and {Narayanan}, Vijay and {Postman}, Marc and {Reichart}, Daniel E. and {Richmond}, Michael and {Schneider}, Donald P. and {Schlegel}, David J. and {Strauss}, Michael A. and {SubbaRao}, Mark and {Tucker}, Douglas L. and {Vanden Berk}, Daniel and {Vogeley}, Michael S. and {Weinberg}, David H. and {Yanny}, Brian},
        title = "{Spectroscopic Target Selection for the Sloan Digital Sky Survey: The Luminous Red Galaxy Sample}",
      journal = {\aj},
         year = 2001,
        month = nov,
       volume = {122},
       number = {5},
        pages = {2267-2280},
          doi = {10.1086/323717},
archivePrefix = {arXiv},
       eprint = {astro-ph/0108153},
 primaryClass = {astro-ph},
       adsurl = {https://ui.adsabs.harvard.edu/abs/2001AJ....122.2267E}
}

@ARTICLE{barber2007,
       author = {{Barber}, Tom and {Meiksin}, Avery and {Murphy}, Tara},
        title = "{Properties of luminous red galaxies in the Sloan Digital Sky Survey}",
      journal = {\mnras},
         year = 2007,
        month = may,
       volume = {377},
       number = {2},
        pages = {787-805},
          doi = {10.1111/j.1365-2966.2007.11650.x},
archivePrefix = {arXiv},
       eprint = {astro-ph/0611053},
 primaryClass = {astro-ph},
       adsurl = {https://ui.adsabs.harvard.edu/abs/2007MNRAS.377..787B}
}

@ARTICLE{pillepich2018,
       author = {{Pillepich}, Annalisa and {Springel}, Volker and {Nelson}, Dylan and {Genel}, Shy and {Naiman}, Jill and {Pakmor}, R{\"u}diger and {Hernquist}, Lars and {Torrey}, Paul and {Vogelsberger}, Mark and {Weinberger}, Rainer and {Marinacci}, Federico},
        title = "{Simulating galaxy formation with the IllustrisTNG model}",
      journal = {\mnras},
         year = 2018,
        month = jan,
       volume = {473},
       number = {3},
        pages = {4077-4106},
          doi = {10.1093/mnras/stx2656},
archivePrefix = {arXiv},
       eprint = {1703.02970},
 primaryClass = {astro-ph.GA},
       adsurl = {https://ui.adsabs.harvard.edu/abs/2018MNRAS.473.4077P}
}

@BOOK{osterbrock1989,
       author = {{Osterbrock}, Donald E.},
        title = "{Astrophysics of gaseous nebulae and active galactic nuclei}",
         year = 1989,
       adsurl = {https://ui.adsabs.harvard.edu/abs/1989agna.book.....O}
}

@ARTICLE{donnari2019,
       author = {{Donnari}, Martina and {Pillepich}, Annalisa and {Nelson}, Dylan and {Vogelsberger}, Mark and {Genel}, Shy and {Weinberger}, Rainer and {Marinacci}, Federico and {Springel}, Volker and {Hernquist}, Lars},
        title = "{The star formation activity of IllustrisTNG galaxies: main sequence, UVJ diagram, quenched fractions, and systematics}",
      journal = {\mnras},
         year = 2019,
        month = jun,
       volume = {485},
       number = {4},
        pages = {4817-4840},
          doi = {10.1093/mnras/stz712},
archivePrefix = {arXiv},
       eprint = {1812.07584},
 primaryClass = {astro-ph.GA},
       adsurl = {https://ui.adsabs.harvard.edu/abs/2019MNRAS.485.4817D}
}

@ARTICLE{donnari2021,
       author = {{Donnari}, Martina and {Pillepich}, Annalisa and {Nelson}, Dylan and {Marinacci}, Federico and {Vogelsberger}, Mark and {Hernquist}, Lars},
        title = "{Quenched fractions in the IllustrisTNG simulations: comparison with observations and other theoretical models}",
      journal = {\mnras},
         year = 2021,
        month = oct,
       volume = {506},
       number = {4},
        pages = {4760-4780},
          doi = {10.1093/mnras/stab1950},
archivePrefix = {arXiv},
       eprint = {2008.00004},
 primaryClass = {astro-ph.GA},
       adsurl = {https://ui.adsabs.harvard.edu/abs/2021MNRAS.506.4760D}
}

@ARTICLE{springel2018,
       author = {{Springel}, Volker and {Pakmor}, R{\"u}diger and {Pillepich}, Annalisa and {Weinberger}, Rainer and {Nelson}, Dylan and {Hernquist}, Lars and {Vogelsberger}, Mark and {Genel}, Shy and {Torrey}, Paul and {Marinacci}, Federico and {Naiman}, Jill},
        title = "{First results from the IllustrisTNG simulations: matter and galaxy clustering}",
      journal = {\mnras},
         year = 2018,
        month = mar,
       volume = {475},
       number = {1},
        pages = {676-698},
          doi = {10.1093/mnras/stx3304},
archivePrefix = {arXiv},
       eprint = {1707.03397},
 primaryClass = {astro-ph.GA},
       adsurl = {https://ui.adsabs.harvard.edu/abs/2018MNRAS.475..676S}
}

@ARTICLE{planck2016,
       author = {{Planck Collaboration} and {Ade}, P.~A.~R. and {Aghanim}, N. and {Arnaud}, M. and {Ashdown}, M. and {Aumont}, J. and {Baccigalupi}, C. and {Banday}, A.~J. and {Barreiro}, R.~B. and {Bartlett}, J.~G. and {Bartolo}, N. and {Battaner}, E. and {Battye}, R. and {Benabed}, K. and {Beno{\^\i}t}, A. and {Benoit-L{\'e}vy}, A. and {Bernard}, J. -P. and {Bersanelli}, M. and {Bielewicz}, P. and {Bock}, J.~J. and {Bonaldi}, A. and {Bonavera}, L. and {Bond}, J.~R. and {Borrill}, J. and {Bouchet}, F.~R. and {Boulanger}, F. and {Bucher}, M. and {Burigana}, C. and {Butler}, R.~C. and {Calabrese}, E. and {Cardoso}, J. -F. and {Catalano}, A. and {Challinor}, A. and {Chamballu}, A. and {Chary}, R. -R. and {Chiang}, H.~C. and {Chluba}, J. and {Christensen}, P.~R. and {Church}, S. and {Clements}, D.~L. and {Colombi}, S. and {Colombo}, L.~P.~L. and {Combet}, C. and {Coulais}, A. and {Crill}, B.~P. and {Curto}, A. and {Cuttaia}, F. and {Danese}, L. and {Davies}, R.~D. and {Davis}, R.~J. and {de Bernardis}, P. and {de Rosa}, A. and {de Zotti}, G. and {Delabrouille}, J. and {D{\'e}sert}, F. -X. and {Di Valentino}, E. and {Dickinson}, C. and {Diego}, J.~M. and {Dolag}, K. and {Dole}, H. and {Donzelli}, S. and {Dor{\'e}}, O. and {Douspis}, M. and {Ducout}, A. and {Dunkley}, J. and {Dupac}, X. and {Efstathiou}, G. and {Elsner}, F. and {En{\ss}lin}, T.~A. and {Eriksen}, H.~K. and {Farhang}, M. and {Fergusson}, J. and {Finelli}, F. and {Forni}, O. and {Frailis}, M. and {Fraisse}, A.~A. and {Franceschi}, E. and {Frejsel}, A. and {Galeotta}, S. and {Galli}, S. and {Ganga}, K. and {Gauthier}, C. and {Gerbino}, M. and {Ghosh}, T. and {Giard}, M. and {Giraud-H{\'e}raud}, Y. and {Giusarma}, E. and {Gjerl{\o}w}, E. and {Gonz{\'a}lez-Nuevo}, J. and {G{\'o}rski}, K.~M. and {Gratton}, S. and {Gregorio}, A. and {Gruppuso}, A. and {Gudmundsson}, J.~E. and {Hamann}, J. and {Hansen}, F.~K. and {Hanson}, D. and {Harrison}, D.~L. and {Helou}, G. and {Henrot-Versill{\'e}}, S. and {Hern{\'a}ndez-Monteagudo}, C. and {Herranz}, D. and {Hildebrandt}, S.~R. and {Hivon}, E. and {Hobson}, M. and {Holmes}, W.~A. and {Hornstrup}, A. and {Hovest}, W. and {Huang}, Z. and {Huffenberger}, K.~M. and {Hurier}, G. and {Jaffe}, A.~H. and {Jaffe}, T.~R. and {Jones}, W.~C. and {Juvela}, M. and {Keih{\"a}nen}, E. and {Keskitalo}, R. and {Kisner}, T.~S. and {Kneissl}, R. and {Knoche}, J. and {Knox}, L. and {Kunz}, M. and {Kurki-Suonio}, H. and {Lagache}, G. and {L{\"a}hteenm{\"a}ki}, A. and {Lamarre}, J. -M. and {Lasenby}, A. and {Lattanzi}, M. and {Lawrence}, C.~R. and {Leahy}, J.~P. and {Leonardi}, R. and {Lesgourgues}, J. and {Levrier}, F. and {Lewis}, A. and {Liguori}, M. and {Lilje}, P.~B. and {Linden-V{\o}rnle}, M. and {L{\'o}pez-Caniego}, M. and {Lubin}, P.~M. and {Mac{\'\i}as-P{\'e}rez}, J.~F. and {Maggio}, G. and {Maino}, D. and {Mandolesi}, N. and {Mangilli}, A. and {Marchini}, A. and {Maris}, M. and {Martin}, P.~G. and {Martinelli}, M. and {Mart{\'\i}nez-Gonz{\'a}lez}, E. and {Masi}, S. and {Matarrese}, S. and {McGehee}, P. and {Meinhold}, P.~R. and {Melchiorri}, A. and {Melin}, J. -B. and {Mendes}, L. and {Mennella}, A. and {Migliaccio}, M. and {Millea}, M. and {Mitra}, S. and {Miville-Desch{\^e}nes}, M. -A. and {Moneti}, A. and {Montier}, L. and {Morgante}, G. and {Mortlock}, D. and {Moss}, A. and {Munshi}, D. and {Murphy}, J.~A. and {Naselsky}, P. and {Nati}, F. and {Natoli}, P. and {Netterfield}, C.~B. and {N{\o}rgaard-Nielsen}, H.~U. and {Noviello}, F. and {Novikov}, D. and {Novikov}, I. and {Oxborrow}, C.~A. and {Paci}, F. and {Pagano}, L. and {Pajot}, F. and {Paladini}, R. and {Paoletti}, D. and {Partridge}, B. and {Pasian}, F. and {Patanchon}, G. and {Pearson}, T.~J. and {Perdereau}, O. and {Perotto}, L. and {Perrotta}, F. and {Pettorino}, V. and {Piacentini}, F. and {Piat}, M. and {Pierpaoli}, E. and {Pietrobon}, D. and {Plaszczynski}, S. and {Pointecouteau}, E. and {Polenta}, G. and {Popa}, L. and {Pratt}, G.~W. and {Pr{\'e}zeau}, G.},
        title = "{Planck 2015 results. XIII. Cosmological parameters}",
      journal = {\aap},
         year = 2016,
        month = sep,
       volume = {594},
          eid = {A13},
        pages = {A13},
          doi = {10.1051/0004-6361/201525830},
archivePrefix = {arXiv},
       eprint = {1502.01589},
 primaryClass = {astro-ph.CO},
       adsurl = {https://ui.adsabs.harvard.edu/abs/2016A&A...594A..13P}
}

@ARTICLE{bernardi2010,
       author = {{Bernardi}, M. and {Shankar}, F. and {Hyde}, J.~B. and {Mei}, S. and {Marulli}, F. and {Sheth}, R.~K.},
        title = "{Galaxy luminosities, stellar masses, sizes, velocity dispersions as a function of morphological type}",
      journal = {\mnras},
         year = 2010,
        month = jun,
       volume = {404},
       number = {4},
        pages = {2087-2122},
          doi = {10.1111/j.1365-2966.2010.16425.x},
archivePrefix = {arXiv},
       eprint = {0910.1093},
 primaryClass = {astro-ph.CO},
       adsurl = {https://ui.adsabs.harvard.edu/abs/2010MNRAS.404.2087B}
}

@ARTICLE{longhetti2009,
       author = {{Longhetti}, M. and {Saracco}, P.},
        title = "{Stellar mass estimates in early-type galaxies: procedures, uncertainties and models dependence}",
      journal = {\mnras},
         year = 2009,
        month = apr,
       volume = {394},
       number = {2},
        pages = {774-794},
          doi = {10.1111/j.1365-2966.2008.14375.x},
archivePrefix = {arXiv},
       eprint = {0811.4041},
 primaryClass = {astro-ph},
       adsurl = {https://ui.adsabs.harvard.edu/abs/2009MNRAS.394..774L}
}

@ARTICLE{reid2016,
       author = {{Reid}, Beth and {Ho}, Shirley and {Padmanabhan}, Nikhil and {Percival}, Will J. and {Tinker}, Jeremy and {Tojeiro}, Rita and {White}, Martin and {Eisenstein}, Daniel J. and {Maraston}, Claudia and {Ross}, Ashley J. and {S{\'a}nchez}, Ariel G. and {Schlegel}, David and {Sheldon}, Erin and {Strauss}, Michael A. and {Thomas}, Daniel and {Wake}, David and {Beutler}, Florian and {Bizyaev}, Dmitry and {Bolton}, Adam S. and {Brownstein}, Joel R. and {Chuang}, Chia-Hsun and {Dawson}, Kyle and {Harding}, Paul and {Kitaura}, Francisco-Shu and {Leauthaud}, Alexie and {Masters}, Karen and {McBride}, Cameron K. and {More}, Surhud and {Olmstead}, Matthew D. and {Oravetz}, Daniel and {Nuza}, Sebasti{\'a}n E. and {Pan}, Kaike and {Parejko}, John and {Pforr}, Janine and {Prada}, Francisco and {Rodr{\'\i}guez-Torres}, Sergio and {Salazar-Albornoz}, Salvador and {Samushia}, Lado and {Schneider}, Donald P. and {Sc{\'o}ccola}, Claudia G. and {Simmons}, Audrey and {Vargas-Magana}, Mariana},
        title = "{SDSS-III Baryon Oscillation Spectroscopic Survey Data Release 12: galaxy target selection and large-scale structure catalogues}",
      journal = {\mnras},
         year = 2016,
        month = jan,
       volume = {455},
       number = {2},
        pages = {1553-1573},
          doi = {10.1093/mnras/stv2382},
archivePrefix = {arXiv},
       eprint = {1509.06529},
 primaryClass = {astro-ph.CO},
       adsurl = {https://ui.adsabs.harvard.edu/abs/2016MNRAS.455.1553R}
}

@ARTICLE{desi2016b,
       author = {{DESI Collaboration} and {Aghamousa}, Amir and {Aguilar}, Jessica and {Ahlen}, Steve and {Alam}, Shadab and {Allen}, Lori E. and {Allende Prieto}, Carlos and {Annis}, James and {Bailey}, Stephen and {Balland}, Christophe and {Ballester}, Otger and {Baltay}, Charles and {Beaufore}, Lucas and {Bebek}, Chris and {Beers}, Timothy C. and {Bell}, Eric F. and {Bernal}, Jos{\'e} Luis and {Besuner}, Robert and {Beutler}, Florian and {Blake}, Chris and {Bleuler}, Hannes and {Blomqvist}, Michael and {Blum}, Robert and {Bolton}, Adam S. and {Briceno}, Cesar and {Brooks}, David and {Brownstein}, Joel R. and {Buckley-Geer}, Elizabeth and {Burden}, Angela and {Burtin}, Etienne and {Busca}, Nicolas G. and {Cahn}, Robert N. and {Cai}, Yan-Chuan and {Cardiel-Sas}, Laia and {Carlberg}, Raymond G. and {Carton}, Pierre-Henri and {Casas}, Ricard and {Castander}, Francisco J. and {Cervantes-Cota}, Jorge L. and {Claybaugh}, Todd M. and {Close}, Madeline and {Coker}, Carl T. and {Cole}, Shaun and {Comparat}, Johan and {Cooper}, Andrew P. and {Cousinou}, M. -C. and {Crocce}, Martin and {Cuby}, Jean-Gabriel and {Cunningham}, Daniel P. and {Davis}, Tamara M. and {Dawson}, Kyle S. and {de la Macorra}, Axel and {De Vicente}, Juan and {Delubac}, Timoth{\'e}e and {Derwent}, Mark and {Dey}, Arjun and {Dhungana}, Govinda and {Ding}, Zhejie and {Doel}, Peter and {Duan}, Yutong T. and {Ealet}, Anne and {Edelstein}, Jerry and {Eftekharzadeh}, Sarah and {Eisenstein}, Daniel J. and {Elliott}, Ann and {Escoffier}, St{\'e}phanie and {Evatt}, Matthew and {Fagrelius}, Parker and {Fan}, Xiaohui and {Fanning}, Kevin and {Farahi}, Arya and {Farihi}, Jay and {Favole}, Ginevra and {Feng}, Yu and {Fernandez}, Enrique and {Findlay}, Joseph R. and {Finkbeiner}, Douglas P. and {Fitzpatrick}, Michael J. and {Flaugher}, Brenna and {Flender}, Samuel and {Font-Ribera}, Andreu and {Forero-Romero}, Jaime E. and {Fosalba}, Pablo and {Frenk}, Carlos S. and {Fumagalli}, Michele and {Gaensicke}, Boris T. and {Gallo}, Giuseppe and {Garcia-Bellido}, Juan and {Gaztanaga}, Enrique and {Pietro Gentile Fusillo}, Nicola and {Gerard}, Terry and {Gershkovich}, Irena and {Giannantonio}, Tommaso and {Gillet}, Denis and {Gonzalez-de-Rivera}, Guillermo and {Gonzalez-Perez}, Violeta and {Gott}, Shelby and {Graur}, Or and {Gutierrez}, Gaston and {Guy}, Julien and {Habib}, Salman and {Heetderks}, Henry and {Heetderks}, Ian and {Heitmann}, Katrin and {Hellwing}, Wojciech A. and {Herrera}, David A. and {Ho}, Shirley and {Holland}, Stephen and {Honscheid}, Klaus and {Huff}, Eric and {Hutchinson}, Timothy A. and {Huterer}, Dragan and {Hwang}, Ho Seong and {Illa Laguna}, Joseph Maria and {Ishikawa}, Yuzo and {Jacobs}, Dianna and {Jeffrey}, Niall and {Jelinsky}, Patrick and {Jennings}, Elise and {Jiang}, Linhua and {Jimenez}, Jorge and {Johnson}, Jennifer and {Joyce}, Richard and {Jullo}, Eric and {Juneau}, St{\'e}phanie and {Kama}, Sami and {Karcher}, Armin and {Karkar}, Sonia and {Kehoe}, Robert and {Kennamer}, Noble and {Kent}, Stephen and {Kilbinger}, Martin and {Kim}, Alex G. and {Kirkby}, David and {Kisner}, Theodore and {Kitanidis}, Ellie and {Kneib}, Jean-Paul and {Koposov}, Sergey and {Kovacs}, Eve and {Koyama}, Kazuya and {Kremin}, Anthony and {Kron}, Richard and {Kronig}, Luzius and {Kueter-Young}, Andrea and {Lacey}, Cedric G. and {Lafever}, Robin and {Lahav}, Ofer and {Lambert}, Andrew and {Lampton}, Michael and {Landriau}, Martin and {Lang}, Dustin and {Lauer}, Tod R. and {Le Goff}, Jean-Marc and {Le Guillou}, Laurent and {Le Van Suu}, Auguste and {Lee}, Jae Hyeon and {Lee}, Su-Jeong and {Leitner}, Daniela and {Lesser}, Michael and {Levi}, Michael E. and {L'Huillier}, Benjamin and {Li}, Baojiu and {Liang}, Ming and {Lin}, Huan and {Linder}, Eric and {Loebman}, Sarah R. and {Luki{\'c}}, Zarija and {Ma}, Jun and {MacCrann}, Niall and {Magneville}, Christophe and {Makarem}, Laleh and {Manera}, Marc and {Manser}, Christopher J. and {Marshall}, Robert and {Martini}, Paul and {Massey}, Richard and {Matheson}, Thomas and {McCauley}, Jeremy and {McDonald}, Patrick and {McGreer}, Ian D. and {Meisner}, Aaron and {Metcalfe}, Nigel and {Miller}, Timothy N. and {Miquel}, Ramon and {Moustakas}, John and {Myers}, Adam and {Naik}, Milind and {Newman}, Jeffrey A. and {Nichol}, Robert C. and {Nicola}, Andrina and {Nicolati da Costa}, Luiz and {Nie}, Jundan and {Niz}, Gustavo and {Norberg}, Peder and {Nord}, Brian and {Norman}, Dara and {Nugent}, Peter and {O'Brien}, Thomas and {Oh}, Minji and {Olsen}, Knut A.~G.},
        title = "{The DESI Experiment Part II: Instrument Design}",
      journal = {arXiv e-prints},
         year = 2016,
        month = oct,
          eid = {arXiv:1611.00037},
        pages = {arXiv:1611.00037},
          doi = {10.48550/arXiv.1611.00037},
archivePrefix = {arXiv},
       eprint = {1611.00037},
 primaryClass = {astro-ph.IM},
       adsurl = {https://ui.adsabs.harvard.edu/abs/2016arXiv161100037D}
}

@ARTICLE{knowles2023,
       author = {{Knowles}, Adam T. and {Sansom}, A.~E. and {Vazdekis}, A. and {Allende Prieto}, C.},
        title = "{sMILES SSPs: a library of semi-empirical MILES stellar population models with variable [{\ensuremath{\alpha}}/Fe] abundances}",
      journal = {\mnras},
         year = 2023,
        month = aug,
       volume = {523},
       number = {3},
        pages = {3450-3470},
          doi = {10.1093/mnras/stad1647},
archivePrefix = {arXiv},
       eprint = {2306.05942},
 primaryClass = {astro-ph.GA},
       adsurl = {https://ui.adsabs.harvard.edu/abs/2023MNRAS.523.3450K}
}

@ARTICLE{leauthaud2016,
       author = {{Leauthaud}, Alexie and {Bundy}, Kevin and {Saito}, Shun and {Tinker}, Jeremy and {Maraston}, Claudia and {Tojeiro}, Rita and {Huang}, Song and {Brownstein}, Joel R. and {Schneider}, Donald P. and {Thomas}, Daniel},
        title = "{The Stripe 82 Massive Galaxy Project - II. Stellar mass completeness of spectroscopic galaxy samples from the Baryon Oscillation Spectroscopic Survey}",
      journal = {\mnras},
         year = 2016,
        month = apr,
       volume = {457},
       number = {4},
        pages = {4021-4037},
          doi = {10.1093/mnras/stw117},
archivePrefix = {arXiv},
       eprint = {1507.04752},
 primaryClass = {astro-ph.GA},
       adsurl = {https://ui.adsabs.harvard.edu/abs/2016MNRAS.457.4021L}
}

@ARTICLE{werle2022,
       author = {{Werle}, Ariel and {Poggianti}, Bianca and {Moretti}, Alessia and {Bellhouse}, Callum and {Vulcani}, Benedetta and {Gullieuszik}, Marco and {Radovich}, Mario and {Fritz}, Jacopo and {Ignesti}, Alessandro and {Richard}, Johan and {Soucail}, Genevi{\`e}ve and {Bruzual}, Gustavo and {Charlot}, Stephane and {Mingozzi}, Matilde and {Bacchini}, Cecilia and {Tomicic}, Neven and {Smith}, Rory and {Kulier}, Andrea and {Peluso}, Giorgia and {Franchetto}, Andrea},
        title = "{Post-starburst Galaxies in the Centers of Intermediate-redshift Clusters}",
      journal = {\apj},
         year = 2022,
        month = may,
       volume = {930},
       number = {1},
          eid = {43},
        pages = {43},
          doi = {10.3847/1538-4357/ac5f06},
archivePrefix = {arXiv},
       eprint = {2203.08862},
 primaryClass = {astro-ph.GA},
       adsurl = {https://ui.adsabs.harvard.edu/abs/2022ApJ...930...43W}
}

@ARTICLE{zhou2020,
       author = {{Zhou}, Rongpu and {Newman}, Jeffrey A. and {Dawson}, Kyle S. and {Eisenstein}, Daniel J. and {Brooks}, David D. and {Dey}, Arjun and {Dey}, Biprateep and {Duan}, Yutong and {Eftekharzadeh}, Sarah and {Gazta{\~n}aga}, Enrique and {Kehoe}, Robert and {Landriau}, Martin and {Levi}, Michael E. and {Licquia}, Timothy C. and {Meisner}, Aaron M. and {Moustakas}, John and {Myers}, Adam D. and {Palanque-Delabrouille}, Nathalie and {Poppett}, Claire and {Prada}, Francisco and {Raichoor}, Anand and {Schlegel}, David J. and {Schubnell}, Michael and {Staten}, Ryan and {Tarl{\'e}}, Gregory and {Y{\`e}che}, Christophe},
        title = "{Preliminary Target Selection for the DESI Luminous Red Galaxy (LRG) Sample}",
      journal = {Research Notes of the American Astronomical Society},
         year = 2020,
        month = oct,
       volume = {4},
       number = {10},
          eid = {181},
        pages = {181},
          doi = {10.3847/2515-5172/abc0f4},
archivePrefix = {arXiv},
       eprint = {2010.11282},
 primaryClass = {astro-ph.CO},
       adsurl = {https://ui.adsabs.harvard.edu/abs/2020RNAAS...4..181Z}
}

@ARTICLE{zhou2023,
       author = {{Zhou}, Rongpu and {Dey}, Biprateep and {Newman}, Jeffrey A. and {Eisenstein}, Daniel J. and {Dawson}, K. and {Bailey}, S. and {Berti}, A. and {Guy}, J. and {Lan}, Ting-Wen and {Zou}, H. and {Aguilar}, J. and {Ahlen}, S. and {Alam}, Shadab and {Brooks}, D. and {de la Macorra}, A. and {Dey}, A. and {Dhungana}, G. and {Fanning}, K. and {Font-Ribera}, A. and {Gontcho}, S. Gontcho A. and {Honscheid}, K. and {Ishak}, Mustapha and {Kisner}, T. and {Kov{\'a}cs}, A. and {Kremin}, A. and {Landriau}, M. and {Levi}, Michael E. and {Magneville}, C. and {Manera}, Marc and {Martini}, P. and {Meisner}, Aaron M. and {Miquel}, R. and {Moustakas}, J. and {Myers}, Adam D. and {Nie}, Jundan and {Palanque-Delabrouille}, N. and {Percival}, W.~J. and {Poppett}, C. and {Prada}, F. and {Raichoor}, A. and {Ross}, A.~J. and {Schlafly}, E. and {Schlegel}, D. and {Schubnell}, M. and {Tarl{\'e}}, Gregory and {Weaver}, B.~A. and {Wechsler}, R.~H. and {Y{\'e}che}, Christophe and {Zhou}, Zhimin},
        title = "{Target Selection and Validation of DESI Luminous Red Galaxies}",
      journal = {\aj},
         year = 2023,
        month = feb,
       volume = {165},
       number = {2},
          eid = {58},
        pages = {58},
          doi = {10.3847/1538-3881/aca5fb},
archivePrefix = {arXiv},
       eprint = {2208.08515},
 primaryClass = {astro-ph.CO},
       adsurl = {https://ui.adsabs.harvard.edu/abs/2023AJ....165...58Z}
}

@ARTICLE{desidr12025arXiv,
       author = {{DESI Collaboration} and {Abdul-Karim}, M. and {Adame}, A.~G. and {Aguado}, D. and {Aguilar}, J. and {Ahlen}, S. and {Alam}, S. and {Aldering}, G. and {Alexander}, D.~M. and {Alfarsy}, R. and {Allen}, L. and {Allende Prieto}, C. and {Alves}, O. and {Anand}, A. and {Andrade}, U. and {Armengaud}, E. and {Avila}, S. and {Aviles}, A. and {Awan}, H. and {Bailey}, S. and {Baleato Lizancos}, A. and {Ballester}, O. and {Bault}, A. and {Bautista}, J. and {BenZvi}, S. and {Beraldo e Silva}, L. and {Bermejo-Climent}, J.~R. and {Beutler}, F. and {Bianchi}, D. and {Blake}, C. and {Blum}, R. and {Bolton}, A.~S. and {Bonici}, M. and {Brieden}, S. and {Brodzeller}, A. and {Brooks}, D. and {Buckley-Geer}, E. and {Burtin}, E. and {Canning}, R. and {Carnero Rosell}, A. and {Carr}, A. and {Carrilho}, P. and {Casas}, L. and {Castander}, F.~J. and {Cereskaite}, R. and {Cervantes-Cota}, J.~L. and {Chaussidon}, E. and {Chaves-Montero}, J. and {Chen}, S. and {Chen}, X. and {Claybaugh}, T. and {Cole}, S. and {Cooper}, A.~P. and {Cousinou}, M. -C. and {Cuceu}, A. and {Davis}, T.~M. and {Dawson}, K.~S. and {de Belsunce}, R. and {de la Cruz}, R. and {de la Macorra}, A. and {de Mattia}, A. and {Deiosso}, N. and {Della Costa}, J. and {Demina}, R. and {Demirbozan}, U. and {DeRose}, J. and {Dey}, A. and {Dey}, B. and {Ding}, J. and {Ding}, Z. and {Doel}, P. and {Douglass}, K. and {Dowicz}, M. and {Ebina}, H. and {Edelstein}, J. and {Eisenstein}, D.~J. and {Elbers}, W. and {Emas}, N. and {Escoffier}, S. and {Fagrelius}, P. and {Fan}, X. and {Fanning}, K. and {Fawcett}, V.~A. and {Fern\textbackslash'andez-Garc\textbackslash'ia}, E. and {Ferraro}, S. and {Findlay}, N. and {Font-Ribera}, A. and {Forero-Romero}, J.~E. and {Forero-S\textbackslash'anchez}, D. and {Frenk}, C.~S. and {G\textbackslash''ansicke}, B.~T. and {Galbany}, L. and {Garc\textbackslash'ia-Bellido}, J. and {Garcia-Quintero}, C. and {Garrison}, L.~H. and {Gazta\textbackslash\raisebox{-0.5ex}}, E. and {Gil-Mar\textbackslash'in}, H. and {Gnedin}, O.~Y. and {Gontcho}, S. Gontcho A and {Gonzalez-Morales}, A.~X. and {Gonzalez-Perez}, V. and {Gordon}, C. and {Graur}, O. and {Green}, D. and {Gruen}, D. and {Gsponer}, R. and {Guandalin}, C. and {Gutierrez}, G. and {Guy}, J. and {Hahn}, C. and {Han}, J.~J. and {Han}, J. and {He}, S. and {Herrera-Alcantar}, H.~K. and {Honscheid}, K. and {Hou}, J. and {Howlett}, C. and {Huterer}, D. and {Ir\textbackslashv\{s\}i\textbackslashv\{c\}}, V. and {Ishak}, M. and {Jacques}, A. and {Jimenez}, J. and {Jing}, Y.~P. and {Joachimi}, B. and {Joudaki}, S. and {Joyce}, R. and {Jullo}, E. and {Juneau}, S. and {Kara\textbackslashc\{c\}ayl\{\textbackslashi\}}, N.~G. and {Karim}, T. and {Kehoe}, R. and {Kent}, S. and {Khederlarian}, A. and {Kirkby}, D. and {Kisner}, T. and {Kitaura}, F. -S. and {Kizhuprakkat}, N. and {Kong}, H. and {Koposov}, S.~E. and {Kremin}, A. and {Krolewski}, A. and {Lahav}, O. and {Lai}, Y. and {Lamman}, C. and {Lan}, T. -W. and {Landriau}, M. and {Lang}, D. and {Lange}, J.~U. and {Lasker}, J. and {Le Goff}, J.~M. and {Le Guillou}, L. and {Leauthaud}, A. and {Levi}, M.~E. and {Li}, S. and {Li}, T.~S. and {Lodha}, K. and {Lokken}, M. and {Luo}, Y. and {Magneville}, C. and {Manera}, M. and {Manser}, C.~J. and {Margala}, D. and {Martini}, P. and {Maus}, M. and {McCullough}, J. and {McDonald}, P. and {Medina}, G.~E. and {Medina-Varela}, L. and {Meisner}, A. and {Mena-Fern\textbackslash'andez}, J. and {Menegas}, A. and {Mezcua}, M. and {Miquel}, R. and {Montero-Camacho}, P. and {Moon}, J. and {Moustakas}, J. and {Mu\textbackslash\raisebox{-0.5ex}tildenoz-Guti\textbackslash'errez}, A. and {Mu\textbackslash\raisebox{-0.5ex}tildenoz-Santos}, D. and {Myers}, A.~D. and {Myles}, J. and {Nadathur}, S. and {Najita}, J. and {Napolitano}, L. and {Newman}, J.~A. and {Nikakhtar}, F. and {Nikutta}, R. and {Niz}, G. and {Noriega}, H.~E. and {Padmanabhan}, N. and {Paillas}, E. and {Palanque-Delabrouille}, N. and {Palmese}, A. and {Pan}, J. and {Pan}, Z. and {Parkinson}, D. and {Peacock}, J. and {Percival}, W.~J. and {P\textbackslash'erez-Fern\textbackslash'andez}, A. and {P\textbackslash'erez-R\textbackslash`afols}, I. and {Peterson}, P.},
        title = "{Data Release 1 of the Dark Energy Spectroscopic Instrument}",
      journal = {arXiv e-prints},
         year = 2025,
        month = mar,
          eid = {arXiv:2503.14745},
        pages = {arXiv:2503.14745},
          doi = {10.48550/arXiv.2503.14745},
archivePrefix = {arXiv},
       eprint = {2503.14745},
 primaryClass = {astro-ph.CO},
       adsurl = {https://ui.adsabs.harvard.edu/abs/2025arXiv250314745D}
}

@ARTICLE{maraston2013,
       author = {{Maraston}, Claudia and {Pforr}, Janine and {Henriques}, Bruno M. and {Thomas}, Daniel and {Wake}, David and {Brownstein}, Joel R. and {Capozzi}, Diego and {Tinker}, Jeremy and {Bundy}, Kevin and {Skibba}, Ramin A. and {Beifiori}, Alessandra and {Nichol}, Robert C. and {Edmondson}, Edd and {Schneider}, Donald P. and {Chen}, Yanmei and {Masters}, Karen L. and {Steele}, Oliver and {Bolton}, Adam S. and {York}, Donald G. and {Weaver}, Benjamin A. and {Higgs}, Tim and {Bizyaev}, Dmitry and {Brewington}, Howard and {Malanushenko}, Elena and {Malanushenko}, Viktor and {Snedden}, Stephanie and {Oravetz}, Daniel and {Pan}, Kaike and {Shelden}, Alaina and {Simmons}, Audrey},
        title = "{Stellar masses of SDSS-III/BOSS galaxies at z {\ensuremath{\sim}} 0.5 and constraints to galaxy formation models}",
      journal = {\mnras},
         year = 2013,
        month = nov,
       volume = {435},
       number = {4},
        pages = {2764-2792},
          doi = {10.1093/mnras/stt1424},
archivePrefix = {arXiv},
       eprint = {1207.6114},
 primaryClass = {astro-ph.CO},
       adsurl = {https://ui.adsabs.harvard.edu/abs/2013MNRAS.435.2764M}
}

@ARTICLE{eisenstein2011,
       author = {{Eisenstein}, Daniel J. and {Weinberg}, David H. and {Agol}, Eric and {Aihara}, Hiroaki and {Allende Prieto}, Carlos and {Anderson}, Scott F. and {Arns}, James A. and {Aubourg}, {\'E}ric and {Bailey}, Stephen and {Balbinot}, Eduardo and {Barkhouser}, Robert and {Beers}, Timothy C. and {Berlind}, Andreas A. and {Bickerton}, Steven J. and {Bizyaev}, Dmitry and {Blanton}, Michael R. and {Bochanski}, John J. and {Bolton}, Adam S. and {Bosman}, Casey T. and {Bovy}, Jo and {Brandt}, W.~N. and {Breslauer}, Ben and {Brewington}, Howard J. and {Brinkmann}, J. and {Brown}, Peter J. and {Brownstein}, Joel R. and {Burger}, Dan and {Busca}, Nicolas G. and {Campbell}, Heather and {Cargile}, Phillip A. and {Carithers}, William C. and {Carlberg}, Joleen K. and {Carr}, Michael A. and {Chang}, Liang and {Chen}, Yanmei and {Chiappini}, Cristina and {Comparat}, Johan and {Connolly}, Natalia and {Cortes}, Marina and {Croft}, Rupert A.~C. and {Cunha}, Katia and {da Costa}, Luiz N. and {Davenport}, James R.~A. and {Dawson}, Kyle and {De Lee}, Nathan and {Porto de Mello}, Gustavo F. and {de Simoni}, Fernando and {Dean}, Janice and {Dhital}, Saurav and {Ealet}, Anne and {Ebelke}, Garrett L. and {Edmondson}, Edward M. and {Eiting}, Jacob M. and {Escoffier}, Stephanie and {Esposito}, Massimiliano and {Evans}, Michael L. and {Fan}, Xiaohui and {Femen{\'\i}a Castell{\'a}}, Bruno and {Dutra Ferreira}, Leticia and {Fitzgerald}, Greg and {Fleming}, Scott W. and {Font-Ribera}, Andreu and {Ford}, Eric B. and {Frinchaboy}, Peter M. and {Garc{\'\i}a P{\'e}rez}, Ana Elia and {Gaudi}, B. Scott and {Ge}, Jian and {Ghezzi}, Luan and {Gillespie}, Bruce A. and {Gilmore}, G. and {Girardi}, L{\'e}o and {Gott}, J. Richard and {Gould}, Andrew and {Grebel}, Eva K. and {Gunn}, James E. and {Hamilton}, Jean-Christophe and {Harding}, Paul and {Harris}, David W. and {Hawley}, Suzanne L. and {Hearty}, Frederick R. and {Hennawi}, Joseph F. and {Gonz{\'a}lez Hern{\'a}ndez}, Jonay I. and {Ho}, Shirley and {Hogg}, David W. and {Holtzman}, Jon A. and {Honscheid}, Klaus and {Inada}, Naohisa and {Ivans}, Inese I. and {Jiang}, Linhua and {Jiang}, Peng and {Johnson}, Jennifer A. and {Jordan}, Cathy and {Jordan}, Wendell P. and {Kauffmann}, Guinevere and {Kazin}, Eyal and {Kirkby}, David and {Klaene}, Mark A. and {Knapp}, G.~R. and {Kneib}, Jean-Paul and {Kochanek}, C.~S. and {Koesterke}, Lars and {Kollmeier}, Juna A. and {Kron}, Richard G. and {Lampeitl}, Hubert and {Lang}, Dustin and {Lawler}, James E. and {Le Goff}, Jean-Marc and {Lee}, Brian L. and {Lee}, Young Sun and {Leisenring}, Jarron M. and {Lin}, Yen-Ting and {Liu}, Jian and {Long}, Daniel C. and {Loomis}, Craig P. and {Lucatello}, Sara and {Lundgren}, Britt and {Lupton}, Robert H. and {Ma}, Bo and {Ma}, Zhibo and {MacDonald}, Nicholas and {Mack}, Claude and {Mahadevan}, Suvrath and {Maia}, Marcio A.~G. and {Majewski}, Steven R. and {Makler}, Martin and {Malanushenko}, Elena and {Malanushenko}, Viktor and {Mandelbaum}, Rachel and {Maraston}, Claudia and {Margala}, Daniel and {Maseman}, Paul and {Masters}, Karen L. and {McBride}, Cameron K. and {McDonald}, Patrick and {McGreer}, Ian D. and {McMahon}, Richard G. and {Mena Requejo}, Olga and {M{\'e}nard}, Brice and {Miralda-Escud{\'e}}, Jordi and {Morrison}, Heather L. and {Mullally}, Fergal and {Muna}, Demitri and {Murayama}, Hitoshi and {Myers}, Adam D. and {Naugle}, Tracy and {Neto}, Angelo Fausti and {Nguyen}, Duy Cuong and {Nichol}, Robert C. and {Nidever}, David L. and {O'Connell}, Robert W. and {Ogando}, Ricardo L.~C. and {Olmstead}, Matthew D. and {Oravetz}, Daniel J. and {Padmanabhan}, Nikhil and {Paegert}, Martin and {Palanque-Delabrouille}, Nathalie and {Pan}, Kaike and {Pandey}, Parul and {Parejko}, John K. and {P{\^a}ris}, Isabelle and {Pellegrini}, Paulo and {Pepper}, Joshua and {Percival}, Will J. and {Petitjean}, Patrick and {Pfaffenberger}, Robert and {Pforr}, Janine and {Phleps}, Stefanie and {Pichon}, Christophe and {Pieri}, Matthew M. and {Prada}, Francisco and {Price-Whelan}, Adrian M. and {Raddick}, M. Jordan and {Ramos}, Beatriz H.~F. and {Reid}, I. Neill and {Reyle}, Celine and {Rich}, James and {Richards}, Gordon T. and {Rieke}, George H. and {Rieke}, Marcia J. and {Rix}, Hans-Walter and {Robin}, Annie C. and {Rocha-Pinto}, Helio J. and {Rockosi}, Constance M. and {Roe}, Natalie A. and {Rollinde}, Emmanuel and {Ross}, Ashley J. and {Ross}, Nicholas P. and {Rossetto}, Bruno and {S{\'a}nchez}, Ariel G. and {Santiago}, Basilio and {Sayres}, Conor and {Schiavon}, Ricardo and {Schlegel}, David J. and {Schlesinger}, Katharine J. and {Schmidt}, Sarah J. and {Schneider}, Donald P. and {Sellgren}, Kris and {Shelden}, Alaina and {Sheldon}, Erin and {Shetrone}, Matthew},
        title = "{SDSS-III: Massive Spectroscopic Surveys of the Distant Universe, the Milky Way, and Extra-Solar Planetary Systems}",
      journal = {\aj},
         year = 2011,
        month = sep,
       volume = {142},
       number = {3},
          eid = {72},
        pages = {72},
          doi = {10.1088/0004-6256/142/3/72},
archivePrefix = {arXiv},
       eprint = {1101.1529},
 primaryClass = {astro-ph.IM},
       adsurl = {https://ui.adsabs.harvard.edu/abs/2011AJ....142...72E}
}

@ARTICLE{dawson2013,
       author = {{Dawson}, Kyle S. and {Schlegel}, David J. and {Ahn}, Christopher P. and {Anderson}, Scott F. and {Aubourg}, {\'E}ric and {Bailey}, Stephen and {Barkhouser}, Robert H. and {Bautista}, Julian E. and {Beifiori}, Alessandra and {Berlind}, Andreas A. and {Bhardwaj}, Vaishali and {Bizyaev}, Dmitry and {Blake}, Cullen H. and {Blanton}, Michael R. and {Blomqvist}, Michael and {Bolton}, Adam S. and {Borde}, Arnaud and {Bovy}, Jo and {Brandt}, W.~N. and {Brewington}, Howard and {Brinkmann}, Jon and {Brown}, Peter J. and {Brownstein}, Joel R. and {Bundy}, Kevin and {Busca}, N.~G. and {Carithers}, William and {Carnero}, Aurelio R. and {Carr}, Michael A. and {Chen}, Yanmei and {Comparat}, Johan and {Connolly}, Natalia and {Cope}, Frances and {Croft}, Rupert A.~C. and {Cuesta}, Antonio J. and {da Costa}, Luiz N. and {Davenport}, James R.~A. and {Delubac}, Timoth{\'e}e and {de Putter}, Roland and {Dhital}, Saurav and {Ealet}, Anne and {Ebelke}, Garrett L. and {Eisenstein}, Daniel J. and {Escoffier}, S. and {Fan}, Xiaohui and {Filiz Ak}, N. and {Finley}, Hayley and {Font-Ribera}, Andreu and {G{\'e}nova-Santos}, R. and {Gunn}, James E. and {Guo}, Hong and {Haggard}, Daryl and {Hall}, Patrick B. and {Hamilton}, Jean-Christophe and {Harris}, Ben and {Harris}, David W. and {Ho}, Shirley and {Hogg}, David W. and {Holder}, Diana and {Honscheid}, Klaus and {Huehnerhoff}, Joe and {Jordan}, Beatrice and {Jordan}, Wendell P. and {Kauffmann}, Guinevere and {Kazin}, Eyal A. and {Kirkby}, David and {Klaene}, Mark A. and {Kneib}, Jean-Paul and {Le Goff}, Jean-Marc and {Lee}, Khee-Gan and {Long}, Daniel C. and {Loomis}, Craig P. and {Lundgren}, Britt and {Lupton}, Robert H. and {Maia}, Marcio A.~G. and {Makler}, Martin and {Malanushenko}, Elena and {Malanushenko}, Viktor and {Mandelbaum}, Rachel and {Manera}, Marc and {Maraston}, Claudia and {Margala}, Daniel and {Masters}, Karen L. and {McBride}, Cameron K. and {McDonald}, Patrick and {McGreer}, Ian D. and {McMahon}, Richard G. and {Mena}, Olga and {Miralda-Escud{\'e}}, Jordi and {Montero-Dorta}, Antonio D. and {Montesano}, Francesco and {Muna}, Demitri and {Myers}, Adam D. and {Naugle}, Tracy and {Nichol}, Robert C. and {Noterdaeme}, Pasquier and {Nuza}, Sebasti{\'a}n E. and {Olmstead}, Matthew D. and {Oravetz}, Audrey and {Oravetz}, Daniel J. and {Owen}, Russell and {Padmanabhan}, Nikhil and {Palanque-Delabrouille}, Nathalie and {Pan}, Kaike and {Parejko}, John K. and {P{\^a}ris}, Isabelle and {Percival}, Will J. and {P{\'e}rez-Fournon}, Ismael and {P{\'e}rez-R{\`a}fols}, Ignasi and {Petitjean}, Patrick and {Pfaffenberger}, Robert and {Pforr}, Janine and {Pieri}, Matthew M. and {Prada}, Francisco and {Price-Whelan}, Adrian M. and {Raddick}, M. Jordan and {Rebolo}, Rafael and {Rich}, James and {Richards}, Gordon T. and {Rockosi}, Constance M. and {Roe}, Natalie A. and {Ross}, Ashley J. and {Ross}, Nicholas P. and {Rossi}, Graziano and {Rubi{\~n}o-Martin}, J.~A. and {Samushia}, Lado and {S{\'a}nchez}, Ariel G. and {Sayres}, Conor and {Schmidt}, Sarah J. and {Schneider}, Donald P. and {Sc{\'o}ccola}, C.~G. and {Seo}, Hee-Jong and {Shelden}, Alaina and {Sheldon}, Erin and {Shen}, Yue and {Shu}, Yiping and {Slosar}, An{\v{z}}e and {Smee}, Stephen A. and {Snedden}, Stephanie A. and {Stauffer}, Fritz and {Steele}, Oliver and {Strauss}, Michael A. and {Streblyanska}, Alina and {Suzuki}, Nao and {Swanson}, Molly E.~C. and {Tal}, Tomer and {Tanaka}, Masayuki and {Thomas}, Daniel and {Tinker}, Jeremy L. and {Tojeiro}, Rita and {Tremonti}, Christy A. and {Vargas Maga{\~n}a}, M. and {Verde}, Licia and {Viel}, Matteo and {Wake}, David A. and {Watson}, Mike and {Weaver}, Benjamin A. and {Weinberg}, David H. and {Weiner}, Benjamin J. and {West}, Andrew A. and {White}, Martin and {Wood-Vasey}, W.~M. and {Yeche}, Christophe and {Zehavi}, Idit and {Zhao}, Gong-Bo and {Zheng}, Zheng},
        title = "{The Baryon Oscillation Spectroscopic Survey of SDSS-III}",
      journal = {\aj},
         year = 2013,
        month = jan,
       volume = {145},
       number = {1},
          eid = {10},
        pages = {10},
          doi = {10.1088/0004-6256/145/1/10},
archivePrefix = {arXiv},
       eprint = {1208.0022},
 primaryClass = {astro-ph.CO},
       adsurl = {https://ui.adsabs.harvard.edu/abs/2013AJ....145...10D}
}

@ARTICLE{ditrani2025,
       author = {{Ditrani}, F.~R. and {Longhetti}, M. and {Iovino}, A. and {Fossati}, M. and {Zhou}, S. and {Bardelli}, S. and {Bolzonella}, M. and {Cucciati}, O. and {Finoguenov}, A. and {Pozzetti}, L. and {Salvato}, M. and {Scodeggio}, M. and {Tasca}, L. and {Vergani}, D. and {Zucca}, E.},
        title = "{The COSMOS Wall at z {\ensuremath{\sim}} 0.73: Quiescent galaxies and their evolution in different environments}",
      journal = {\aap},
         year = 2025,
        month = apr,
       volume = {696},
          eid = {A116},
        pages = {A116},
          doi = {10.1051/0004-6361/202453620},
archivePrefix = {arXiv},
       eprint = {2503.19974},
 primaryClass = {astro-ph.GA},
       adsurl = {https://ui.adsabs.harvard.edu/abs/2025A&A...696A.116D}
}

@ARTICLE{kroupa2001,
       author = {{Kroupa}, Pavel},
        title = "{On the variation of the initial mass function}",
      journal = {\mnras},
         year = 2001,
        month = apr,
       volume = {322},
       number = {2},
        pages = {231-246},
          doi = {10.1046/j.1365-8711.2001.04022.x},
archivePrefix = {arXiv},
       eprint = {astro-ph/0009005},
 primaryClass = {astro-ph},
       adsurl = {https://ui.adsabs.harvard.edu/abs/2001MNRAS.322..231K}
}

@ARTICLE{scoville2007,
       author = {{Scoville}, N. and {Aussel}, H. and {Brusa}, M. and {Capak}, P. and {Carollo}, C.~M. and {Elvis}, M. and {Giavalisco}, M. and {Guzzo}, L. and {Hasinger}, G. and {Impey}, C. and {Kneib}, J. -P. and {LeFevre}, O. and {Lilly}, S.~J. and {Mobasher}, B. and {Renzini}, A. and {Rich}, R.~M. and {Sanders}, D.~B. and {Schinnerer}, E. and {Schminovich}, D. and {Shopbell}, P. and {Taniguchi}, Y. and {Tyson}, N.~D.},
        title = "{The Cosmic Evolution Survey (COSMOS): Overview}",
      journal = {\apjs},
         year = 2007,
        month = sep,
       volume = {172},
       number = {1},
        pages = {1-8},
          doi = {10.1086/516585},
archivePrefix = {arXiv},
       eprint = {astro-ph/0612305},
 primaryClass = {astro-ph},
       adsurl = {https://ui.adsabs.harvard.edu/abs/2007ApJS..172....1S}
}

@ARTICLE{muzzin2013,
       author = {{Muzzin}, Adam and {Marchesini}, Danilo and {Stefanon}, Mauro and {Franx}, Marijn and {McCracken}, Henry J. and {Milvang-Jensen}, Bo and {Dunlop}, James S. and {Fynbo}, J.~P.~U. and {Brammer}, Gabriel and {Labb{\'e}}, Ivo and {van Dokkum}, Pieter G.},
        title = "{The Evolution of the Stellar Mass Functions of Star-forming and Quiescent Galaxies to z = 4 from the COSMOS/UltraVISTA Survey}",
      journal = {\apj},
         year = 2013,
        month = nov,
       volume = {777},
       number = {1},
          eid = {18},
        pages = {18},
          doi = {10.1088/0004-637X/777/1/18},
archivePrefix = {arXiv},
       eprint = {1303.4409},
 primaryClass = {astro-ph.CO},
       adsurl = {https://ui.adsabs.harvard.edu/abs/2013ApJ...777...18M}
}

@ARTICLE{bevacqua2023,
       author = {{Bevacqua}, Davide and {Saracco}, Paolo and {La Barbera}, Francesco and {D'Ago}, Giuseppe and {De Propris}, Roberto and {Ferreras}, Ignacio and {Gallazzi}, Anna and {Pasquali}, Anna and {Spiniello}, Chiara},
        title = "{The elemental abundance of quiescent galaxies in the LEGA-C survey: the (non-)evolution of [{\ensuremath{\alpha}}/Fe] from z = 0.75 to z = 0}",
      journal = {\mnras},
         year = 2023,
        month = nov,
       volume = {525},
       number = {3},
        pages = {4219-4230},
          doi = {10.1093/mnras/stad2403},
archivePrefix = {arXiv},
       eprint = {2308.03441},
 primaryClass = {astro-ph.GA},
       adsurl = {https://ui.adsabs.harvard.edu/abs/2023MNRAS.525.4219B}
}

@ARTICLE{thomas2010,
       author = {{Thomas}, Daniel and {Maraston}, Claudia and {Schawinski}, Kevin and {Sarzi}, Marc and {Silk}, Joseph},
        title = "{Environment and self-regulation in galaxy formation}",
      journal = {\mnras},
         year = 2010,
        month = jun,
       volume = {404},
       number = {4},
        pages = {1775-1789},
          doi = {10.1111/j.1365-2966.2010.16427.x},
archivePrefix = {arXiv},
       eprint = {0912.0259},
 primaryClass = {astro-ph.CO},
       adsurl = {https://ui.adsabs.harvard.edu/abs/2010MNRAS.404.1775T}
}

@ARTICLE{gallazzi2014,
       author = {{Gallazzi}, Anna and {Bell}, Eric F. and {Zibetti}, Stefano and {Brinchmann}, Jarle and {Kelson}, Daniel D.},
        title = "{Charting the Evolution of the Ages and Metallicities of Massive Galaxies since z = 0.7}",
      journal = {\apj},
         year = 2014,
        month = jun,
       volume = {788},
       number = {1},
          eid = {72},
        pages = {72},
          doi = {10.1088/0004-637X/788/1/72},
archivePrefix = {arXiv},
       eprint = {1404.5624},
 primaryClass = {astro-ph.GA},
       adsurl = {https://ui.adsabs.harvard.edu/abs/2014ApJ...788...72G}
}

@ARTICLE{kuntschner2004,
       author = {{Kuntschner}, H.},
        title = "{Line-of-sight velocity distribution corrections for Lick/IDS indices of early-type galaxies}",
      journal = {\aap},
         year = 2004,
        month = nov,
       volume = {426},
        pages = {737-745},
          doi = {10.1051/0004-6361:20041414},
archivePrefix = {arXiv},
       eprint = {astro-ph/0407219},
 primaryClass = {astro-ph},
       adsurl = {https://ui.adsabs.harvard.edu/abs/2004A&A...426..737K}
}

@ARTICLE{kim2023,
       author = {{Kim}, Jee-Ho and {Belli}, Sirio and {Weinberger}, Rainer},
        title = "{The stellar chemical abundances of simulated massive galaxies at z = 2}",
      journal = {\mnras},
         year = 2023,
        month = jul,
       volume = {523},
       number = {1},
        pages = {849-864},
          doi = {10.1093/mnras/stad1338},
archivePrefix = {arXiv},
       eprint = {2210.14235},
 primaryClass = {astro-ph.GA},
       adsurl = {https://ui.adsabs.harvard.edu/abs/2023MNRAS.523..849K}
}

@ARTICLE{jin2024,
       author = {{Jin}, Shoko and {Trager}, Scott C. and {Dalton}, Gavin B. and {Aguerri}, J. Alfonso L. and {Drew}, J.~E. and {Falc{\'o}n-Barroso}, Jes{\'u}s and {G{\"a}nsicke}, Boris T. and {Hill}, Vanessa and {Iovino}, Angela and {Pieri}, Matthew M. and {Poggianti}, Bianca M. and {Smith}, D.~J.~B. and {Vallenari}, Antonella and {Abrams}, Don Carlos and {Aguado}, David S. and {Antoja}, Teresa and {Arag{\'o}n-Salamanca}, Alfonso and {Ascasibar}, Yago and {Babusiaux}, Carine and {Balcells}, Marc and {Barrena}, R. and {Battaglia}, Giuseppina and {Belokurov}, Vasily and {Bensby}, Thomas and {Bonifacio}, Piercarlo and {Bragaglia}, Angela and {Carrasco}, Esperanza and {Carrera}, Ricardo and {Cornwell}, Daniel J. and {Dom{\'\i}nguez-Palmero}, Lilian and {Duncan}, Kenneth J. and {Famaey}, Benoit and {Fari{\~n}a}, Cecilia and {Gonzalez}, Oscar A. and {Guest}, Steve and {Hatch}, Nina A. and {Hess}, Kelley M. and {Hoskin}, Matthew J. and {Irwin}, Mike and {Knapen}, Johan H. and {Koposov}, Sergey E. and {Kuchner}, Ulrike and {Laigle}, Clotilde and {Lewis}, Jim and {Longhetti}, Marcella and {Lucatello}, Sara and {M{\'e}ndez-Abreu}, Jairo and {Mercurio}, Amata and {Molaeinezhad}, Alireza and {Mongui{\'o}}, Maria and {Morrison}, Sean and {Murphy}, David N.~A. and {Peralta de Arriba}, Luis and {P{\'e}rez}, Isabel and {P{\'e}rez-R{\`a}fols}, Ignasi and {Pic{\'o}}, Sergio and {Raddi}, Roberto and {Romero-G{\'o}mez}, Merc{\`e} and {Royer}, Fr{\'e}d{\'e}ric and {Siebert}, Arnaud and {Seabroke}, George M. and {Som}, Debopam and {Terrett}, David and {Thomas}, Guillaume and {Wesson}, Roger and {Worley}, C. Clare and {Alfaro}, Emilio J. and {Allende Prieto}, Carlos and {Alonso-Santiago}, Javier and {Amos}, Nicholas J. and {Ashley}, Richard P. and {Balaguer-N{\'u}{\~n}ez}, Lola and {Balbinot}, Eduardo and {Bellazzini}, Michele and {Benn}, Chris R. and {Berlanas}, Sara R. and {Bernard}, Edouard J. and {Best}, Philip and {Bettoni}, Daniela and {Bianco}, Andrea and {Bishop}, Georgia and {Blomqvist}, Michael and {Boeche}, Corrado and {Bolzonella}, Micol and {Bonoli}, Silvia and {Bosma}, Albert and {Britavskiy}, Nikolay and {Busarello}, Gianni and {Caffau}, Elisabetta and {Cantat-Gaudin}, Tristan and {Castro-Ginard}, Alfred and {Couto}, Guilherme and {Carbajo-Hijarrubia}, Juan and {Carter}, David and {Casamiquela}, Laia and {Conrado}, Ana M. and {Corcho-Caballero}, Pablo and {Costantin}, Luca and {Deason}, Alis and {de Burgos}, Abel and {De Grandi}, Sabrina and {Di Matteo}, Paola and {Dom{\'\i}nguez-G{\'o}mez}, Jes{\'u}s and {Dorda}, Ricardo and {Drake}, Alyssa and {Dutta}, Rajeshwari and {Erkal}, Denis and {Feltzing}, Sofia and {Ferr{\'e}-Mateu}, Anna and {Feuillet}, Diane and {Figueras}, Francesca and {Fossati}, Matteo and {Franciosini}, Elena and {Frasca}, Antonio and {Fumagalli}, Michele and {Gallazzi}, Anna and {Garc{\'\i}a-Benito}, Rub{\'e}n and {Gentile Fusillo}, Nicola and {Gebran}, Marwan and {Gilbert}, James and {Gledhill}, T.~M. and {Gonz{\'a}lez Delgado}, Rosa M. and {Greimel}, Robert and {Guarcello}, Mario Giuseppe and {Guerra}, Jose and {Gullieuszik}, Marco and {Haines}, Christopher P. and {Hardcastle}, Martin J. and {Harris}, Amy and {Haywood}, Misha and {Helmi}, Amina and {Hernandez}, Nauzet and {Herrero}, Artemio and {Hughes}, Sarah and {Ir{\v{s}}i{\v{c}}}, Vid and {Jablonka}, Pascale and {Jarvis}, Matt J. and {Jordi}, Carme and {Kondapally}, Rohit and {Kordopatis}, Georges and {Krogager}, Jens-Kristian and {La Barbera}, Francesco and {Lam}, Man I. and {Larsen}, S{\o}ren S. and {Lemasle}, Bertrand and {Lewis}, Ian J. and {Lhom{\'e}}, Emilie and {Lind}, Karin and {Lodi}, Marcello and {Longobardi}, Alessia and {Lonoce}, Ilaria and {Magrini}, Laura and {Ma{\'\i}z Apell{\'a}niz}, Jes{\'u}s and {Marchal}, Olivier and {Marco}, Amparo and {Martin}, Nicolas F. and {Matsuno}, Tadafumi and {Maurogordato}, Sophie and {Merluzzi}, Paola and {Miralda-Escud{\'e}}, Jordi and {Molinari}, Emilio and {Monari}, Giacomo and {Morelli}, Lorenzo and {Mottram}, Christopher J. and {Naylor}, Tim and {Negueruela}, Ignacio and {O{\~n}orbe}, Jose and {Pancino}, Elena and {Peirani}, S{\'e}bastien and {Peletier}, Reynier F. and {Pozzetti}, Lucia and {Rainer}, Monica and {Ramos}, Pau and {Read}, Shaun C. and {Rossi}, Elena Maria and {R{\"o}ttgering}, Huub J.~A. and {Rubi{\~n}o-Mart{\'\i}n}, Jose Alberto and {Sabater}, Jose and {San Juan}, Jos{\'e} and {Sanna}, Nicoletta and {Schallig}, Ellen and {Schiavon}, Ricardo P. and {Schultheis}, Mathias and {Serra}, Paolo and {Shimwell}, Timothy W. and {Sim{\'o}n-D{\'\i}az}, Sergio and {Smith}, Russell J. and {Sordo}, Rosanna and {Sorini}, Daniele and {Soubiran}, Caroline and {Starkenburg}, Else and {Steele}, Iain A. and {Stott}, John and {Stuik}, Remko and {Tolstoy}, Eline and {Tortora}, Crescenzo and {Tsantaki}, Maria and {Van der Swaelmen}, Mathieu and {van Weeren}, Reinout J. and {Vergani}, Daniela},
        title = "{The wide-field, multiplexed, spectroscopic facility WEAVE: Survey design, overview, and simulated implementation}",
      journal = {\mnras},
         year = 2024,
        month = may,
       volume = {530},
       number = {3},
        pages = {2688-2730},
          doi = {10.1093/mnras/stad557},
archivePrefix = {arXiv},
       eprint = {2212.03981},
 primaryClass = {astro-ph.IM},
       adsurl = {https://ui.adsabs.harvard.edu/abs/2024MNRAS.530.2688J}
}

@ARTICLE{dejong2019,
       author = {{de Jong}, R.~S. and {Agertz}, O. and {Berbel}, A.~A. and {Aird}, J. and {Alexander}, D.~A. and {Amarsi}, A. and {Anders}, F. and {Andrae}, R. and {Ansarinejad}, B. and {Ansorge}, W. and {Antilogus}, P. and {Anwand-Heerwart}, H. and {Arentsen}, A. and {Arnadottir}, A. and {Asplund}, M. and {Auger}, M. and {Azais}, N. and {Baade}, D. and {Baker}, G. and {Baker}, S. and {Balbinot}, E. and {Baldry}, I.~K. and {Banerji}, M. and {Barden}, S. and {Barklem}, P. and {Barth{\'e}l{\'e}my-Mazot}, E. and {Battistini}, C. and {Bauer}, S. and {Bell}, C.~P.~M. and {Bellido-Tirado}, O. and {Bellstedt}, S. and {Belokurov}, V. and {Bensby}, T. and {Bergemann}, M. and {Bestenlehner}, J.~M. and {Bielby}, R. and {Bilicki}, M. and {Blake}, C. and {Bland-Hawthorn}, J. and {Boeche}, C. and {Boland}, W. and {Boller}, T. and {Bongard}, S. and {Bongiorno}, A. and {Bonifacio}, P. and {Boudon}, D. and {Brooks}, D. and {Brown}, M.~J.~I. and {Brown}, R. and {Br{\"u}ggen}, M. and {Brynnel}, J. and {Brzeski}, J. and {Buchert}, T. and {Buschkamp}, P. and {Caffau}, E. and {Caillier}, P. and {Carrick}, J. and {Casagrande}, L. and {Case}, S. and {Casey}, A. and {Cesarini}, I. and {Cescutti}, G. and {Chapuis}, D. and {Chiappini}, C. and {Childress}, M. and {Christlieb}, N. and {Church}, R. and {Cioni}, M.-R.~L. and {Cluver}, M. and {Colless}, M. and {Collett}, T. and {Comparat}, J. and {Cooper}, A. and {Couch}, W. and {Courbin}, F. and {Croom}, S. and {Croton}, D. and {Daguis{\'e}}, E. and {Dalton}, G. and {Davies}, L.~J.~M. and {Davis}, T. and {de Laverny}, P. and {Deason}, A. and {Dionies}, F. and {Disseau}, K. and {Doel}, P. and {D{\"o}scher}, D. and {Driver}, S.~P. and {Dwelly}, T. and {Eckert}, D. and {Edge}, A. and {Edvardsson}, B. and {Youssoufi}, D.~E. and {Elhaddad}, A. and {Enke}, H. and {Erfanianfar}, G. and {Farrell}, T. and {Fechner}, T. and {Feiz}, C. and {Feltzing}, S. and {Ferreras}, I. and {Feuerstein}, D. and {Feuillet}, D. and {Finoguenov}, A. and {Ford}, D. and {Fotopoulou}, S. and {Fouesneau}, M. and {Frenk}, C. and {Frey}, S. and {Gaessler}, W. and {Geier}, S. and {Gentile Fusillo}, N. and {Gerhard}, O. and {Giannantonio}, T. and {Giannone}, D. and {Gibson}, B. and {Gillingham}, P. and {Gonz{\'a}lez-Fern{\'a}ndez}, C. and {Gonzalez-Solares}, E. and {Gottloeber}, S. and {Gould}, A. and {Grebel}, E.~K. and {Gueguen}, A. and {Guiglion}, G. and {Haehnelt}, M. and {Hahn}, T. and {Hansen}, C.~J. and {Hartman}, H. and {Hauptner}, K. and {Hawkins}, K. and {Haynes}, D. and {Haynes}, R. and {Heiter}, U. and {Helmi}, A. and {Aguayo}, C.~H. and {Hewett}, P. and {Hinton}, S. and {Hobbs}, D. and {Hoenig}, S. and {Hofman}, D. and {Hook}, I. and {Hopgood}, J. and {Hopkins}, A. and {Hourihane}, A. and {Howes}, L. and {Howlett}, C. and {Huet}, T. and {Irwin}, M. and {Iwert}, O. and {Jablonka}, P. and {Jahn}, T. and {Jahnke}, K. and {Jarno}, A. and {Jin}, S. and {Jofre}, P. and {Johl}, D. and {Jones}, D. and {J{\"o}nsson}, H. and {Jordan}, C. and {Karovicova}, I. and {Khalatyan}, A. and {Kelz}, A. and {Kennicutt}, R. and {King}, D. and {Kitaura}, F. and {Klar}, J. and {Klauser}, U. and {Kneib}, J.-P. and {Koch}, A. and {Koposov}, S. and {Kordopatis}, G. and {Korn}, A. and {Kosmalski}, J. and {Kotak}, R. and {Kovalev}, M. and {Kreckel}, K. and {Kripak}, Y. and {Krumpe}, M. and {Kuijken}, K. and {Kunder}, A. and {Kushniruk}, I. and {Lam}, M.~I. and {Lamer}, G. and {Laurent}, F. and {Lawrence}, J. and {Lehmitz}, M. and {Lemasle}, B. and {Lewis}, J. and {Li}, B. and {Lidman}, C. and {Lind}, K. and {Liske}, J. and {Lizon}, J.-L. and {Loveday}, J. and {Ludwig}, H.-G. and {McDermid}, R.~M. and {Maguire}, K. and {Mainieri}, V. and {Mali}, S. and {Mandel}, H.},
        title = "{4MOST: Project overview and information for the First Call for Proposals}",
      journal = {The Messenger},
         year = 2019,
        month = mar,
       volume = {175},
        pages = {3-11},
          doi = {10.18727/0722-6691/5117},
archivePrefix = {arXiv},
       eprint = {1903.02464},
 primaryClass = {astro-ph.IM},
       adsurl = {https://ui.adsabs.harvard.edu/abs/2019Msngr.175....3D}
}

@ARTICLE{labarbera2025,
       author = {{La Barbera}, F. and {Vazdekis}, A. and {Matteucci}, F. and {Spitoni}, E. and {Pasquali}, A. and {Mart{\'\i}n-Navarro}, I.},
        title = "{Elemental abundance ratios for the bulge of M31}",
      journal = {arXiv e-prints},
         year = 2025,
        month = nov,
          eid = {arXiv:2511.15415},
        pages = {arXiv:2511.15415},
          doi = {10.48550/arXiv.2511.15415},
archivePrefix = {arXiv},
       eprint = {2511.15415},
 primaryClass = {astro-ph.GA},
       adsurl = {https://ui.adsabs.harvard.edu/abs/2025arXiv251115415L}
}




\appendix

\section{Consistency between BOSS and DESI}
\label{sec:appconsistency}
\begin{figure*}
   \centering
   \includegraphics[width=0.9\textwidth,height=0.8\textheight,keepaspectratio]{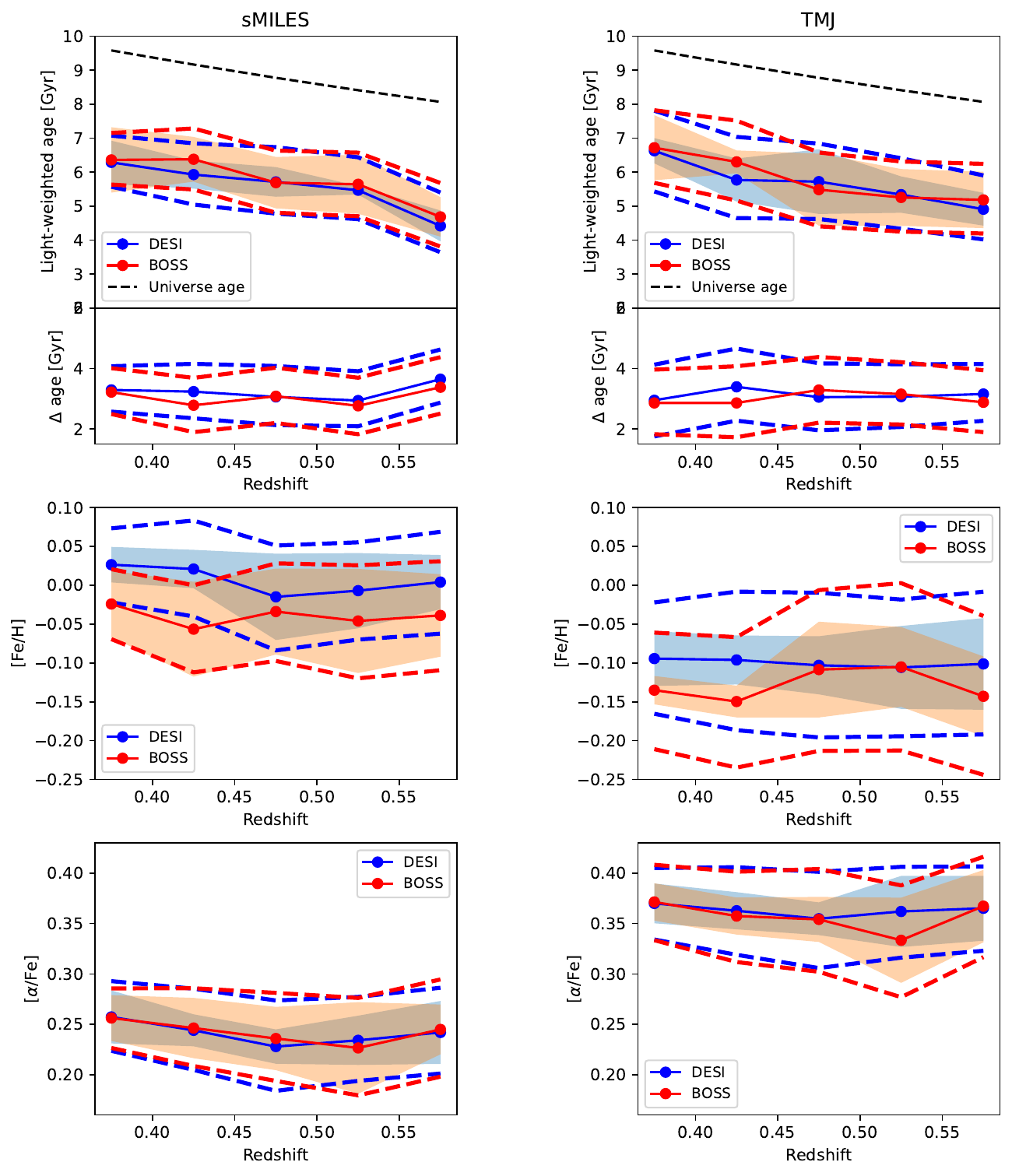}
      \caption{Light-weighted age, [Fe/H] and [$\alpha$/Fe] in the redshift range $0.35<z<0.6$ for the matched DESI and BOSS stacked spectra. Each top panel includes a sub-panel showing the difference between the age of the Universe and the light-weighted age.
      The three panels on the left show the results obtained using sMILES, while the three panels on the right show the corresponding results from TMJ. The data points represent the median value in each redshift bin, the dashed lines cover the typical uncertainties of the stacked spectra, and the shaded regions correspond to the intrinsic scatter within each bin.}
         \label{fig:results_samegalaxies}
   \end{figure*}

We selected LRGs observed in both the BOSS and DESI survey within the redshift range $0.35 < z< 0.6$ (for which we obtained $9854$ galaxies), in order to test the consistency between DESI and BOSS spectra. 
We stacked the individual spectra in groups of $30$ galaxies, using those matched between the BOSS and DESI datasets and we performed the same analysis described in Sect.~\ref{sec:analysis}.
Figure~\ref{fig:results_samegalaxies} shows the results obtained in the measure of light-weighted age, [Fe/H], and [$\alpha$/Fe] for stacks of the matched galaxies in DESI and BOSS, using both sMILES and TMJ models. The results for age and [$\alpha$/Fe] are within $1\sigma$ across the redshift range where BOSS and DESI overlap. However, there is a systematic offset between the [Fe/H] measured from the DESI and BOSS spectra up to $0.05$ dex in both analyses. The systematic offset in stellar metallicity that we found can be explained by the different fibre apertures used by the two instruments. BOSS fibres have a diameter of $2$ arcsec, while DESI fibres are $1.5$ arcsec, implying that BOSS spectra sample a larger fraction of each galaxy. Because massive quiescent galaxies are known to exhibit negative metallicity gradients, typically with
$\Delta[M/H]/\Delta \log R\sim-0.2$ to $-0.4$ \citep[e.g.][]{santucci2020,zibetti2020,parikh2021}, a larger aperture effectively includes a greater contribution from the outer, more metal-poor regions of the galaxies. Assuming a representative gradient of $-0.3$ dex/dex and the relative aperture sizes of DESI and BOSS ($1.5$ arcsec and $2$ arcsec, corresponding to a difference of $\Delta \log \left(\frac{R_{BOSS}}{R_{DESI}}\right) \sim 0.12$), the expected metallicity difference is $\Delta [M/H] \sim -0.04$ dex. This value is fully consistent with the observed offset (
$\sim 0.05$ dex) between the two datasets. This support the interpretation that aperture effects, coupled with intrinsic metallicity gradients, are responsible for the discrepancy.

\section{TNG300-TNG100 comparison}
\label{sec:apptng}
\begin{figure}
   \centering
   \includegraphics[width=0.5\textwidth]{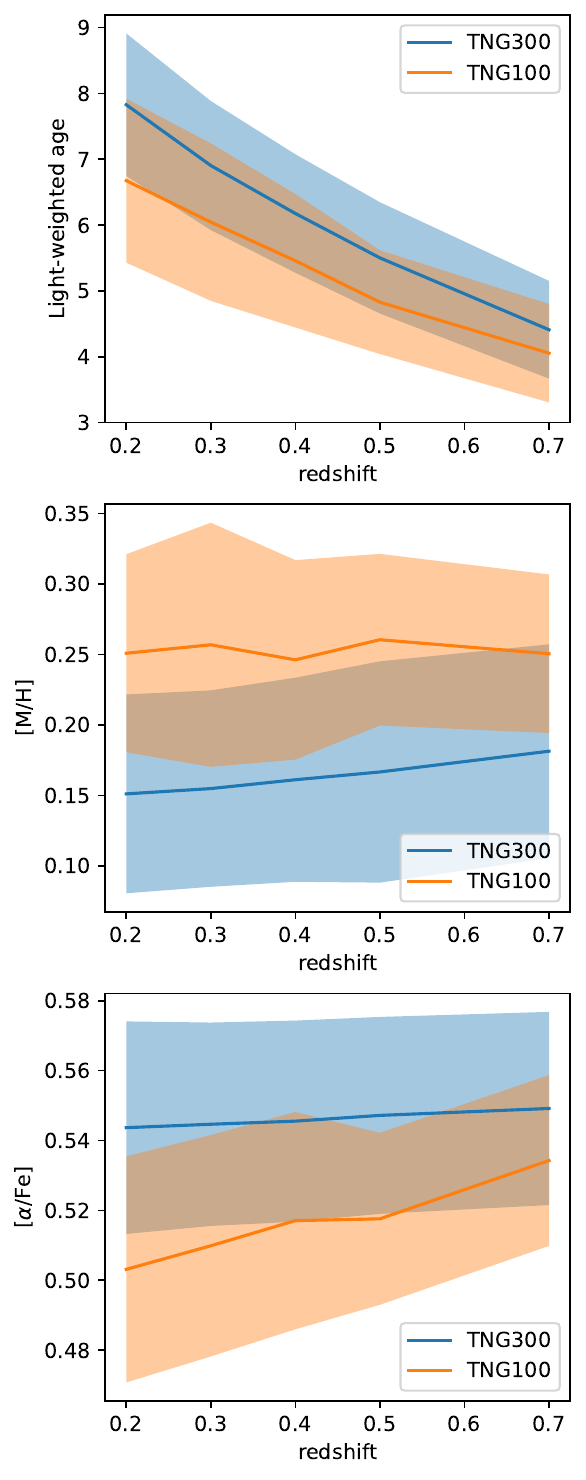}
      \caption{Evolution of the light-weighted age, [M/H] and [$\alpha$/Fe] across the five redshift bins for galaxies in TNG$300$ (blue) and TNG$100$ (orange). The shaded regions represent the dispersion of each parameter within each redshift bin.}
         \label{fig:tng300100}
   \end{figure}

We investigated potential differences of the stellar population parameters of massive quiescent galaxies ($\log (M_*/\mathrm{M_\odot}) > 11.5$ and $\log$(sSFR) $< -11$) within TNG$300$ and TNG$100$ in the $5$ snapshots described in Sect.~\ref{sec:tngcomparison}. Figure ~\ref{fig:tng300100} shows the light-weighted age, stellar metallicity and [$\alpha$/Fe] in the redshift bins $z = [0.2, 0.3, 0.4, 0.5, 0.7]$ for TNG$300$ and TNG$100$. We found that galaxies in TNG$300$ have slightly older ages compared to the ones in TNG$100$, with offset of around $0.5$ Gyr, and well within the dispersion at each redshift. This is consistent with the reported effects of resolution on quenching timescales. Indeed, galaxies in TNG$300$ quench slightly earlier than the ones in TNG$100$, and can explain the slightly older ages \citep[e.g.][]{donnari2019,nelson2019,donnari2021}. For the stellar metallicity, we estimated slightly lower values ($\approx 0.05-0.1$ dex) in galaxies from TNG$300$ in respect to the ones from TNG$100$.

Finally, we found that [$\alpha$/Fe] are marginally higher ($\approx 0.02-0.04$ dex) in galaxies from TNG$300$ than the ones from TNG$100$.
Despite these small differences, the trends and scatter of the stellar population parameters are consistent between the two simulation.

\bsp	
\label{lastpage}
\end{document}